\newif\ifsubmode
\submodefalse

\documentclass[useAMS,usenatbib]{mn2e}

\usepackage{natbib}
\usepackage{epsf}
\usepackage{color}
\usepackage{amsmath}
\usepackage{subfigure}
\usepackage{footnote}
\usepackage{graphicx}
\begin{document}
\title[DYNAMO II: Stellar Kinematics in Two Low Redshift Clumpy Disks]{DYNAMO II: Coupled Stellar and Ionized Gas Kinematics in
  Two Low Redshift Clumpy Disks}
\author[Bassett et al.]{Robert Bassett$^1$,
Karl Glazebrook$^{1,2}$,
David B. Fisher$^1$,
Andrew W. Green$^3$,\newauthor
Emily Wisnioski$^4$,
Danail Obreschkow$^{2,5}$,
Erin Mentuch Cooper$^6$,\newauthor
Roberto G. Abraham$^7$,
Ivana Damjanov$^8$,
Peter J. McGregor$^9$\\
$^1$Centre for Astrophysics and Supercomputing, Swinburne
  University of Technology, P.O. Box 218,\\Hawthorn, VIC 3122,
  Australia\\
$^2$ARC Centre of Excellence for All-sky Astrophysics (CAASTRO)\\
$^3$Australian Astronomical Observatory, P.O. Box 970, North
  Ryde, NSW 1670, Australia\\
$^4$Max Planck Institute for Extraterrestrial Physics,
  Garching, Germany\\
$^5$International Centre for Radio Astronomy Research (ICRAR),
  M468, University of Western Australia,\\35 Stirling Hwy, Crawley, WA
  6009, Australia\\
$^6$Department of Astronomy, The University of Texas at Austin,
Austin, TX 78712, USA\\
$^7$Department of Astronomy \& Astrophysics, University of
  Toronto, 50 St. George St., Toronto,\\ON M5S 3H8, Canada\\
$^8$Harvard-Smithsonian Center for Astrophysics, 60 Garden St.,
  Cambridge, MA 02138\\
$^9$Research School of Astronomy \& Astrophysics, The
  Australian National University, Cotter Rd.,\\Weston, ACT, Australia, 2611}

\pagerange{\pageref{firstpage}--\pageref{lastpage}} \pubyear{2014}

\maketitle

\label{firstpage}

\begin{abstract} 

We study the spatially resolved stellar kinematics of two star-forming 
galaxies at $z\sim0.1$ from the larger DYnamics of Newly Assembled
Massive Objects (DYNAMO) sample. These galaxies, which have been
characterized by high
levels of star formation and large ionized gas velocity dispersions,
are considered possible analogs to high-redshift clumpy disks.
They were observed using the GMOS instrument in integral field
spectroscopy (IFS) mode at the Gemini
Observatory with high spectral resolution (R$\simeq$5400, equivalent to
$\sigma\simeq$24 km s$^{-1}$ at the observed wavelengths) and $\sim 6$ hour
exposure times in order to
measure the resolved stellar kinematics via absorption lines. We also
obtain higher-quality emission line kinematics than previous observations.
The spatial resolution (1.2 kpc) is sufficient to show that the
ionized gas in these
galaxies (as traced by H$\beta$ emission) is morphologically
irregular, forming multiple giant clumps while stellar continuum light
is smooth and well described by an exponential profile. Clumpy
gas morphologies observed in IFS data are confirmed by complementary
narrow band H$\alpha$ imaging from the Hubble Space Telescope. Morphological
differences between the stars and ionized gas
are not reflected dynamically as stellar kinematics are found the be
closely coupled to the kinematics of the ionized gas: both components
are smoothly rotating with large velocity dispersions ($\sim40$ km
s$^{-1}$) suggesting that the high gas dispersions are not primarily
driven by star-formation feedback. In addition, the stellar
population ages of these galaxies are estimated to be quite young
(60-500 Myr). The large velocity dispersions measured for these young
stars suggest that we are seeing the formation of thick disks and/or
stellar bulges in support of recent models which produce these from
clumpy galaxies at high redshift.

\end{abstract}

\begin{keywords}
galaxies: evolution, stars: kinematics and dynamics.
\end{keywords}

\section{Introduction}\label{section:intro}

Understanding the process of galaxy evolution, particularly in its
earliest stages, is a major goal of many research projects in
astrophysics today. Galaxies observed locally are typically 
near the end of their evolution, having
assembled a majority of their stars at high redshift ($z>1$), closer to the peak of
cosmic star formation \citep{madau96,lilly96,hopk06,sobral12}. 

Star-forming
galaxies at high redshift are often found to be composed of a few
large ($\sim1$ kpc) star-forming complexes (``clumps'') with high
molecular gas content (gas fractions of 20-50\%) and high
star-formation efficiencies \citep{daddi10,tacc10,elme11}. These
complexes are more massive (with typical gas masses of
$10^{8}-10^{9}$ $M_{\odot}$) and form stars more efficiently
than similar regions observed in the Milky Way \citep{elme05,forster11,swinbank11,wisnioski12}. 
If star forming regions at high redshift obey local scaling relations \citep{schmidt59,kenni98}, they require high molecular-gas
surface densities to fuel the extreme levels of star formation
observed \citep{guo12}. 

Clumpy, irregular galaxies at high redshift have been observed 
using integral-field spectroscopy (IFS). For studies of galaxy
dynamics, IFS data is used 
to produce detailed two dimensional kinematics maps not possible using traditional
spectroscopic techniques. In recent IFS studies, roughly 1/3
of the high-redshift, clumpy galaxies display the random motions
expected for galaxies undergoing major mergers. The remainder are
either too compact to be reliably classified by the employed
methods, or display ordered rotation characteristic of disk
galaxies. Disk fractions were found to be $20-50\%$ for a number of
samples
\citep{genzel08,forster09,law09,wright09,wisnioski11}. One key
observation of clumpy high-redshift disks are large values of velocity dispersion (as high as $\sim$90
km s$^{-1}$) compared with values of 20-30 km s$^{-1}$ seen locally \citep{andersen06}. \citet{glaz13} provides a comprehensive 
review of kinematic studies at high redshift.

The origin and ultimate fate of the clumps in high redshift galaxies
is a matter of some debate. In some recent models of galaxy formation, massive
high-redshift clumps are shown to spiral inward due to dynamical
friction and coalesce into 
the precursors of local thick disks and galactic bulges
\citep{noguchi99,immeli04,genzel08,dekel09a,cever10,bour14}. 
Alternatively it has been suggested that star formation feedback in the
form of stellar winds and supernovae will cause
clumps to be disrupted before migrating to the proto-galactic
center \citep{murray10,genel12}. These feedback mechanisms, however, are not well
constrained by current observations. In these simulations, the
strength and efficiency of the feedback is known to have significant effects on the evolution
of simulated high-redshift galaxies \citep{mandel13}. Recent observations of
star-forming clumps in high-redshift galaxies have estimated
clump lifetimes ranging from 10 to 200 Myr
\citep{elme09,genzel11,newman12,wuyts12}. 

At high redshift, detailed studies of individual clumpy
galaxies have been primarily based on observations of ionized gas or
broad stellar analysis based on infrared images \citep{wuyts12}. Stellar kinematics
and ages, which must be measured from absorption features, are too
observationally expensive, even using current state of the art
facilities. Such observations, however, could be key to distinguishing
between models involving clump migration and coalescence from extreme
feedback models in which clumps are rapidly dissolved.

If the large velocity dispersions observed from 
H$\alpha$ emission are the result of disruptive feedback, we would
expect to measure stellar velocity
dispersions smaller than those observed for the ionized gas. This would 
indicate a thick gaseous disk, which has been
inflated by strong winds originating from an underlying thinner disk of stars. On the
other hand, a key prediction of the turbulent, clumpy disk model is
$\sigma_{*}=\sigma_{gas}$ which
suggests that the stars are formed from an already turbulent
medium.
Accounting for this initial turbulence may require external
sources of gas \citep[e.g., cold accretion or
galaxy
interactions,][]{keres05,bourn09a,dekel09a,vandevoort11}. These
external sources of gas may drive turbulence directly or through instabilities commonly associated with gas rich
disks \citep{dekel09a,forbes13,bour14}.
As of yet, there is no clear consensus on the true fate of giant
star-forming clumps, however the above two scenarios are currently the
most popular. We note here that in HII regions in the Milky Way,
velocity dispersions of the ionized gas and associated young O, B and
A stars ($\sim$5-10 km s$^{-1}$) are found to be lower than
those of the stellar thin disk ($\sim$20 km s$^{-1}$), with the
largest values found for the oldest stars \citep[stellar thick disk,
$\sim$40 km s$^{-1}$,][]{vdkAA11,pase12}.

While there has been a significant effort to distinguish between
clump migration and feedback dominated disk formation models,
observing the stellar kinematics at high redshift has not been
possible and thus we turn to analogs that we can observed locally. It
is important for analogs for this purpose to have high star-formation
rates, gas
kinematics consistent with turbulent rotating disks, large gas fractions, and clumpy gas
morphologies in order to match key features of high-redshift clumpy disks. 
Studies of low-redshift analogs such as these provide a much
greater level of detail and can be used to quantify the effects of
resolution and sensitivity on high redshift IFS
observations \citep{gonc10,arribas14}. In previous work \citep{green10,green13} we have found a
sample of low-redshift (0.07 $<$ $z$ $<$ 0.14) galaxies with SFRs and
kinematics well matched to high-redshift clumpy galaxies. A subsample
of these have also been shown from observations with the Plateau de
Bure Interferometer to host gas fractions of $\sim$0.4 (Fisher et
al. submitted). 

\begin{table*}
  \caption{Physical Properties}
  \begin{tabular}{ l c c c c c c c c c c c c c }
  \hline\hline
  ID & \textit{z} & L(H$\alpha$)$^{1}$ &
  SFR$_{H\alpha}^{2}$ & $\mathcal{M}_{*}^{3}$ &
 $\sigma_{m,SPIRAL}^{4}$ &
 $\mathcal{M}_{gas}^{5}$ & b/a$^{6}$ & $i^{7}$ &
 $h^{8}$ &
  $n_{b}^{9}$ &
  $h_{b}^{10}$\\
      &  &  (log erg s$^{-1}$) & (M$_{\odot}$yr$^{-1}$) & ($10^{9}$
      M$_{\odot}$) & (km s$^{-1}$) & ($10^{9}$
      M$_{\odot}$) & & (degrees) & kpc & & kpc \\
  \hline
  G 4-1 & 0.129 & 42.36 & 41.61 & 64.74 & 50.2 & 43.68
  & 0.85 & 32.1 & 2.63 & 0.4$\pm$0.6 & 0.2$\pm$0.2
  \\
  G 20-2 & 0.141 & 42.26 & 17.27 & 21.56 & 44.9 &
  19.12 & 0.88 & 28.6 & 2.75 & 0.5$\pm$0.1 & 0.52$\pm$0.06
  \\
  \hline 
  \end{tabular}
  \raggedright 
  $^{1}$The H$\alpha$
    luminosity as measured by previous IFS observations
    \citep{green13}\\
  $^{2}$The star formation rate measured from
    previous IFS observations \citep{green13}\\
  $^{3}$The
    stellar mass reproduced from \citet{kauf03b} scaled by 0.88 to
    convert from their \citet{kroupa01} initial-mass function to that
    of \citet{chab03}\\
  $^{4}$Flux weighted gas velocity dispersion
   from previous IFS observations taken using the SPIRAL IFS\\
  $^{5}$Molecular gas mass inferred from the
   Kennicut-Schmidt relation \citep{green13}\\
  $^{6}$axis ratio
   from 2D surface brightness fitting \citep[GALFIT,][]{peng02} of
   GMOS continuum maps\\
  $^{7}$Inclination calculated from
   axis ratios using the standard equation\\
  $^{8}$scale radius measured from exponential fits to combined
   HST, SDSS, and GMOS data (Section \ref{section:emmap})\\
  $^{9}$S\'{e}rsic index of bulge from fits shown in Figure
    \ref{figure:expprof}\\
  $^{10}$scale radius of bulge from fits shown in Figure \ref{figure:expprof}
  \label{table:physparam}

\end{table*}

In this paper we examine stellar and ionized gas
morphologies of two $z\sim0.1$ disk
galaxies which are found to exhibit both ordered rotation and
turbulent ionized gas from previous IFS observations \citep{green13}. 
Observations considered here are higher in both signal-to-noise ratio (S/N) and spatial resolution
than previous IFS observations of \citet{green13} allowing us to perform
far more detailed morphological and kinematic analysis. We compare
this with
previous IFS observations in order to understand the effects of sensitivity and
resolution on the results of \citep{green13}. Finally, we compare the
resolved stellar and ionized gas kinematics of these galaxies in order to understand the sources of turbulence in young galactic disks.

This paper is laid out as follows: In Section \ref{section:data} we
describe our sample and the observations performed for this study, in Section
\ref{section:analysis} we describe our analysis methodology and
present the results, in Section \ref{section:discussion} we discuss
the implications of our results, and we provide a brief summary in
Section \ref{section:summary}. Throughout this work, we adopt a flat
cosmology with $H_{0}=70$ km s$^{-1}$ Mpc$^{-1}$ and
$\Omega_{M}=0.3$. This results in
a comoving line of sight distance of 533 and 578 Mpc and a spatial
scale of 2.29 kpc/'' and 2.46 kpc/'' for galaxies G 4-1 and G 20-2
respectively. 

\section{Sample Selection and Observations}\label{section:data}

\subsection{{\normalfont DYNAMO} Sample}

Recently, \citet{green10} identified a sample
of low-redshift ($z\simeq.05-0.1$) star-forming galaxies, which are
possible analogs to high redshift clumpy
galaxies. Subsamples of these galaxies have now received extensive multiwavelength
observations as part of the DYnamics of Newly Assembled Massive Objects
(DYNAMO) project. We briefly outline the DYNAMO selection here, however
for full sample-selection details see \citet{green13}. 

All DYNAMO galaxies have been classified as star forming in the
MPA-JHU Value Added Catalog of the Sloan Digital Sky Survey
\citep[SDSS, ][]{york00}. Galaxies hosting active galactic nuclei (AGN)
have been excluded based on their position in the \citet*[BPT,][]{baldw81} diagram. The DYNAMO sample includes 67 galaxies with stellar
masses ranging from $1.09$ to $65.0$ $\times$ $10^{9}$ M$_{\odot}$ and
star-formation rates (SFRs, inferred from H$\alpha$ luminosity of 0.2
to 56.6 M$_{\odot}$ yr$^{-1}$) covering
typical values found on the main sequence of star formation as well as
extremely star-forming galaxies. High H$\alpha$ luminosity galaxies ($>$
$10^{41.5}$ ergs s$^{-1}$) are of
particular interest as possible analogs to high-redshift, clumpy
galaxies. At low redshift however, they are quite rare (the most
likely analogs constitute the upper 95th percentile in H$\alpha$ flux
in SDSS DR4).

DYNAMO galaxies were observed using the
AAOmega-SPIRAL and Wide Field Spectrograph (WiFeS) IFS (0$\farcs$7 and
1$\farcs$0 spatial sampling respectively with typical seeing of 1$\farcs$0-1$\farcs$5). 
Around one-half of DYNAMO galaxies exhibit regular, rotation-like
symmetry. This is comparable to studies at high redshift, such as the
SINS survey, in which roughly one-third are classified as rotators, while the
remainder are either compact, dispersion-dominated systems or mergers
\citep{forster09}. Their properties suggest that highly star-forming, disk-like
galaxies in DYNAMO represent a rare, local sample of young, late-type
systems, which are in an evolutionary state more common at high
redshift ($z\geq1$). 

\subsection{Absorption Line Subsample}

In this paper we focus on two DYNAMO galaxies with
smooth rotation and large velocity dispersions, which are
similar to star-forming disk galaxies seen at high redshift.
Developing accurate dynamical models of these galaxies necessitates
the study of their stellar kinematics. This requires observations of stellar
absorption lines with spectral resolutions of $\simeq$20 km
s$^{-1}$ or better, implying significantly longer exposure times than
needed for ionized gas kinematics. Studies such as these are currently not
possible at high redshift using existing facilities due to the
prohibitively long exposure times required. 

In addition to long exposure times, we also obtain higher spatial
resolution than \citet{green13} to allow for studies of effects such
as beam smearing. Here we present two
galaxies from \citet{green13} which were observed using IFS at the Gemini
Observatory (programs GS-2012B-Q-88 and GN-2012B-Q-130). 
These galaxies were selected based on their disk-like
kinematics and turbulent ionized gas from previous IFS observations
\citep{green13}. The physical properties of these galaxies
(DYNAMO ID's G 4-1 and G 20-2) are summarized in Table \ref{table:physparam}.

\subsection{Observations and Data Reduction}\label{section:dr}

Observations for this study were performed using the
Gemini Multi-Object Spectrograph \citep[GMOS,][]{hook04} in the IFS
mode \citep{alling02}. In this configuration, the
GMOS field of view is comprised of 750 hexagonal fibers, each having a
projected diameter of $0\farcs2$. Of these, 500 fibers are used for target
observations and are arranged in a rectangular
field of view covering $3\farcs5$ $\times$ $5\farcs0$. The remaining 250 fibers are used for
dedicated sky observations taken $1\farcm0$ from the science position. The
detector is made up of three $2048\times4069$ CCD chips separated by
$\sim$0.5 mm gaps.

Observations presented here were performed at the Gemini
South Telescope in October 2011 and January 2012. The
B1200 grating was used which is the highest resolution grating
available (R$\simeq$5400 at the observed wavelengths), giving a dispersion of 0.24 \AA\ per pixel. The
average seeing was measured from stars in our acquisition images using the IRAF
task \textit{psfmeasure} and was found to be $0\farcs5$ (a factor of 3 better
than SPIRAL and WiFeS observations). This corresponds to a physical
distance of 1.15 and 1.23 kpc at the redshifts of G 4-1 and G 20-2
respectively. This is comparable to the SINS-AO sample which has a
typical spatial resolution of 1.6 kpc. Each object was observed at two
positions to maximize the spatial coverage. Exposures in each position were
dithered spatially as well as spectrally, the latter being achieved by
slightly adjusting the grating tilt-angle.

The data was reduced using the Gemini IRAF software package
\citep[version 1.9,][]{turner06}, 
which includes bias subtraction, flat-field correction, fiber tracing
and extraction, wavelength calibration, and sky subtraction. Motivated
by previous studies using similar data \citep{westm07,liu13a}, we created a custom reduction
pipeline that replaces the cosmic ray rejection included in the
standard Gemini-GMOS reduction with the Laplacian edge-detection
algorithm L.A.Cosmic \citep{vandokk01}. This greatly improves the
detection and removal of cosmic rays. This step was performed between
bias subtraction and fiber extraction. 

The data-reduction
process produced an image referred to as a row-stacked spectra (RSS), in which each pixel row
corresponds to the spectrum in a single fiber. The RSS images were then
resampled by linearly interpolating from the center positions of the
hexagonal fibers to create 3D data cubes with $0\farcs1$ square 
pixels. Finally, the individual cubes from each exposure were
spectrally resampled onto the same scale. This is necessary due to a
slight non-uniformity in $\Delta\lambda$ between exposures (which is
attributed to the spectral dithering). The data cubes were then
coadded using a custom IDL script. 

We find a number of clear features in our reduced data cubes.
The spectral range of the final coadded data cubes varies slightly
between our objects, but in the rest-frame of each galaxy the coverage
is the same (by design). For each of our objects, the sampled
rest-frame wavelengths extend from roughly 4000 \AA\ to  5350 \AA. This
wavelength range allows us to measure ionized-gas kinematics using
three Balmer lines (H$\beta\lambda$4861 \AA, H$\gamma\lambda$4340 \AA,
and H$\delta\lambda$4102 \AA) and the [OIII]
doublet at 5007 \AA. Stellar-kinematics measurements are
most strongly constrained by absorption associated with the
same three Balmer transitions seen in emission but some constraint is
also provided by a variety of iron absorption lines throughout
this spectral range.

In Section \ref{section:emmap} we compare morphologies from GMOS IFS
to Hubble Space Telescope (HST) FR647M (Proposal ID 12977, PI Damjanov) imaging in which clumps are well resolved. Our HST observations
include continuum and H$\alpha$ imaging. A full description of this
data will be included in future work (Fisher et al. in prep). This data
illustrates the importance of fully resolving clumps, allowing us to
make robust estimates of clump properties such as size and H$\alpha$
luminosity as well as giving a more reliable measure of clump numbers in our
galaxies. 

\section{Analysis and Results}\label{section:analysis}

\subsection{Mapping Emission: Clumps}\label{section:emmap}

\begin{figure*}
\includegraphics[width=\linewidth]{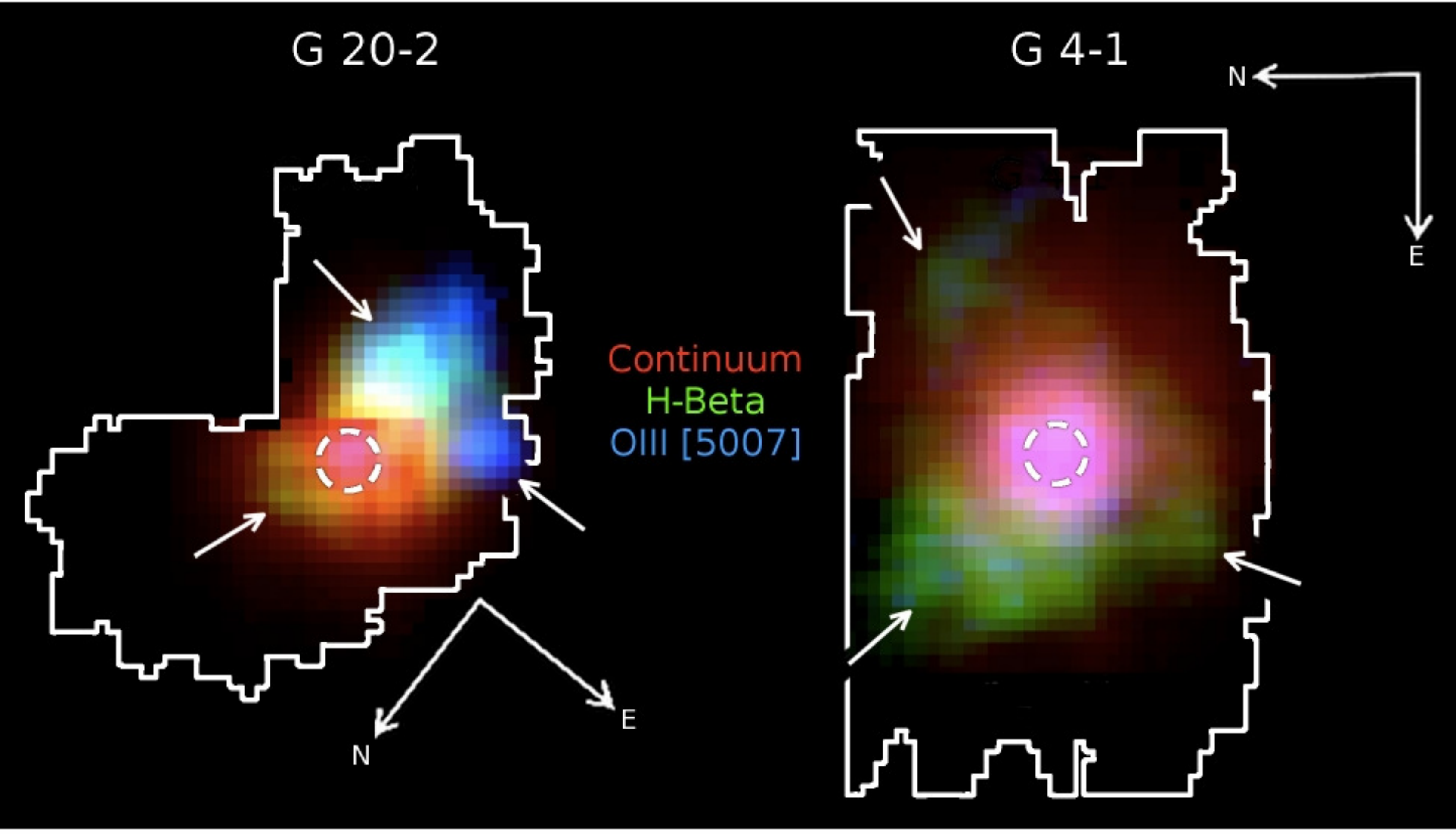}

\vspace{0pt}
 \caption{False color images of our disk galaxies. Red corresponds
   to the median continuum level (rest frame B band) while green and blue show the peak
   flux of the H$\beta$ and [OIII] (5007 \AA) emission lines
   respectively. The color scaling is not the same
   between components, but is chosen to highlight substructure of our
   disk galaxies. The peak of the continuum emission is marked by a
   dashed circle with a diameter equivalent to the average seeing of
   our observations ($\sim$0$\farcs$5$\simeq$ 1.1 kpc at z$=0.14$). In
   both galaxies the continuum peak location is coincident with the
   kinematic center (see Figure \ref{figure:gaskine}). Both
   galaxies exhibit multiple large clumps of gas
   which are offset from the central continuum clump by a distance
   which is approximately equal to the seeing of our observations. The locations of the clumps are
   marked with white arrows. We note the irregularities in
   the central regions of G 20-2 which are due to dead fibers in the detector obscuring the
   central region in multiple exposures. Overplotted are the
   footprints of Figure \ref{figure:gaskine} in white which indicate
   the area over which we can reliably extract ionized gas kinematics
   from our datacubes.}\label{figure:clumps}

\end{figure*}

We use relative
emission-line and continuum strengths to map their baryonic content in order to better understand the structure of our galaxies in comparison with high-redshift clumpy disks.
Figure \ref{figure:clumps} shows false color images of our 
galaxies. Red corresponds to the median continuum
value in each spaxel in the restframe wavelength range of $\sim$3900-4200
\AA, which traces the stellar-mass component. We achieve a continuum
signal to noise ratio of $\sim$15 and $\sim$21 in the
central regions of G 4-1 and G 20-2 respectively. The
ionized gas is traced in green and blue, indicating the peak value of
the H$\beta$ and [OIII] emission lines, respectively. The scaling of
each component was chosen to best 
illustrate the clumpy substructure of our disk galaxies. The white
solid lines correspond to the spatial extent over which we were able to
extract information on ionized gas kinematics. Note the
irregularity in the central regions of G 20-2 are due to dead
fibers in the instrument. Our spatial dithers of 0$\farcs$25 were not
sufficiently large relative to our average seeing ($\sim$0$\farcs$5) to remove this
effect. These dead fibers cause problems
matching the flux between exposures, but do not affect line-profile
shapes and thus will not change our conclusions based on kinematics as
these do not depend on
accurate fluxes.

In both galaxies we observe a single, centrally located peak in
continuum emission, which corresponds spatially with both a peak seen in
\textit{i}-band imaging from SDSS as well as the galaxies' kinematic
centers (see Figures \ref{figure:gaskine} and
\ref{figure:sk202}). We indicate these locations with white dashed circles in Figure
\ref{figure:clumps}, with diameters equivalent to the seeing of our
observations ($\sim$0$\farcs$5, equivalent to 1.15 and 1.23 kpc for G~4-1 and G~20-2 respectively). We also observe multiple peaks in emission-line strength, which are offset
from the continuum peak. We indicate these in Figure \ref{figure:clumps} by
white arrows. The offsets between the continuum peak and emission line
clumps are between 1 and 2 kpc, which is resolved by the seeing of our
observations.

\begin{figure}
\includegraphics[width=\linewidth]{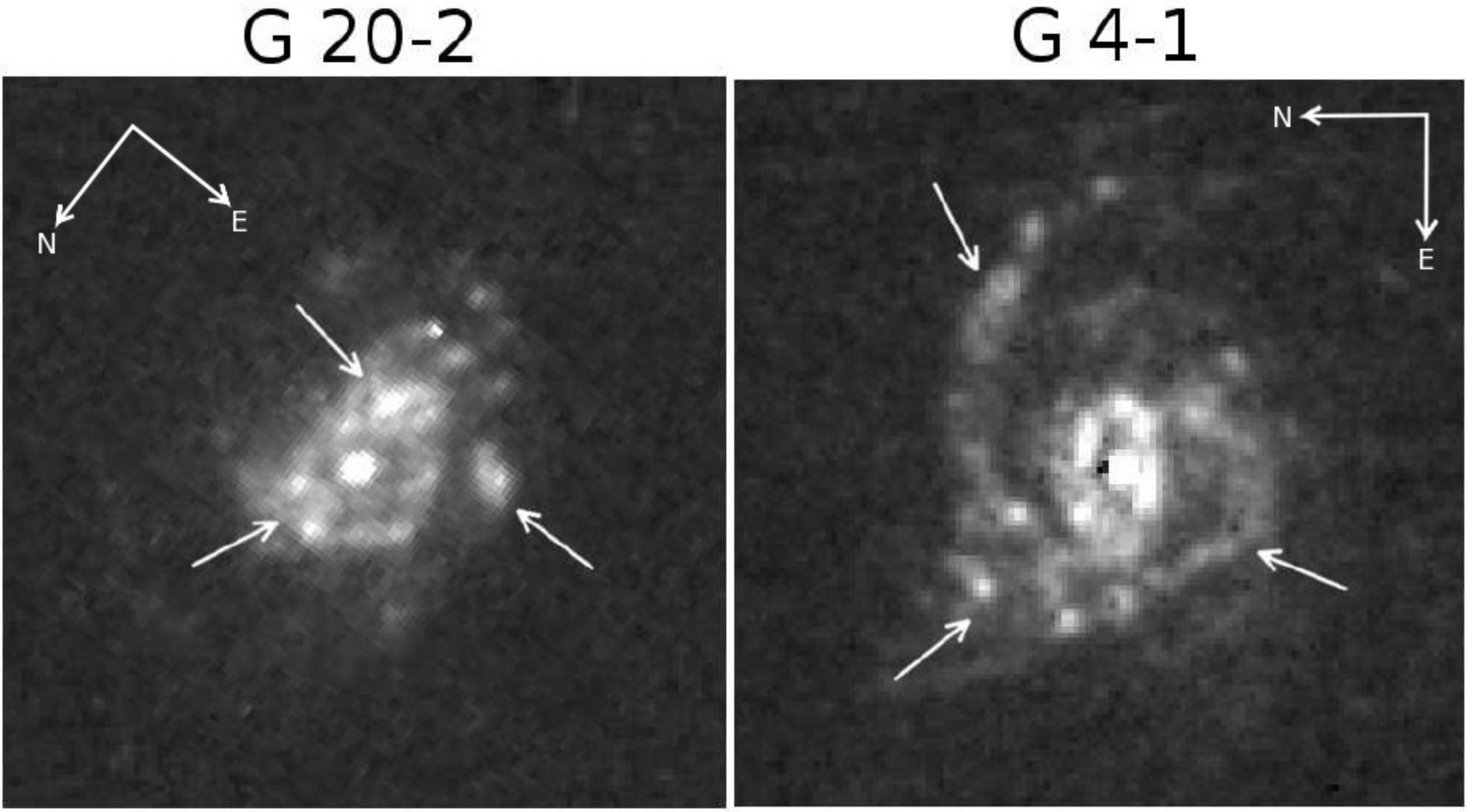}
 \caption{Continuum subtracted H$\alpha$ maps from HST with
   0$\farcs$05 pixel scale. HST resolution of 0$\farcs$1
   corresponds to a physical distance of 229 pc and 246 pc at the redshift
   of G 4-1 and G 20-2 respectively. Comparing with Figure
   \ref{figure:clumps} we find that clumps in the HST imaging are more
   easily distinguished. While some clumps identified from GMOS-IFS
   data are matched to single clumps here, others are found to be
   unresolved collections of multiple clumps. While clump sizes are smaller than would
   be predicted from GMOS-IFS, most clumps are resolved by HST
   providing a lower limit to their sizes of 250 pc. The white
   arrows indicate the same clump locations shown in Figure
   \ref{figure:clumps}. Comparisons between our low redshift IFS data
   and high-resolution, space-based imaging will be valuable for the
   interpretation of similar studies at high redshift.
   } \label{figure:HST}

\end{figure}

Figure \ref{figure:HST} shows continuum subtracted H$\alpha$ maps from
HST with
0$\farcs$05 pixel scale. HST resolution of 0$\farcs$1
corresponds to a physical distance of 229 pc and 246 pc at the redshifts
of G~4-1 and G~20-2 respectively. We identify a number of clumps in the
HST H$\alpha$ maps through visual inspection. These clumps range in luminosity from 10$^{40}$ to
10$^{42}$ ergs~s$^{-1}$ indicating SFRs of $\sim$1-10 M$_{\odot}$
yr$^{-1}$. These numbers are consistent with the properties of clumps
observed in $z\sim$2 disk galaxies \citep{genzel11}. A detailed study
of the clump properties using the HST photometry will be the subject of
future work by the DYNAMO team (Fisher et al. in prep). 

Comparing Figures
\ref{figure:clumps} and \ref{figure:HST} we find that clumps in the HST imaging are more
easily distinguished. Some clumps identified from GMOS-IFS
data correspond to single clumps from HST while others are 
unresolved collections of multiple clumps. Clump sizes are smaller
than would 
be indicated from GMOS-IFS, however, most clumps are well resolved by HST
giving a lower limit to their sizes of 250 pc (HST resolution
at $z=0.13$). This shows that the comparison between our low-redshift
IFS data and high-resolution, space-based imaging will be valuable for
the interpretation of similar studies at high redshift.

\begin{figure}

\includegraphics[width=\linewidth]{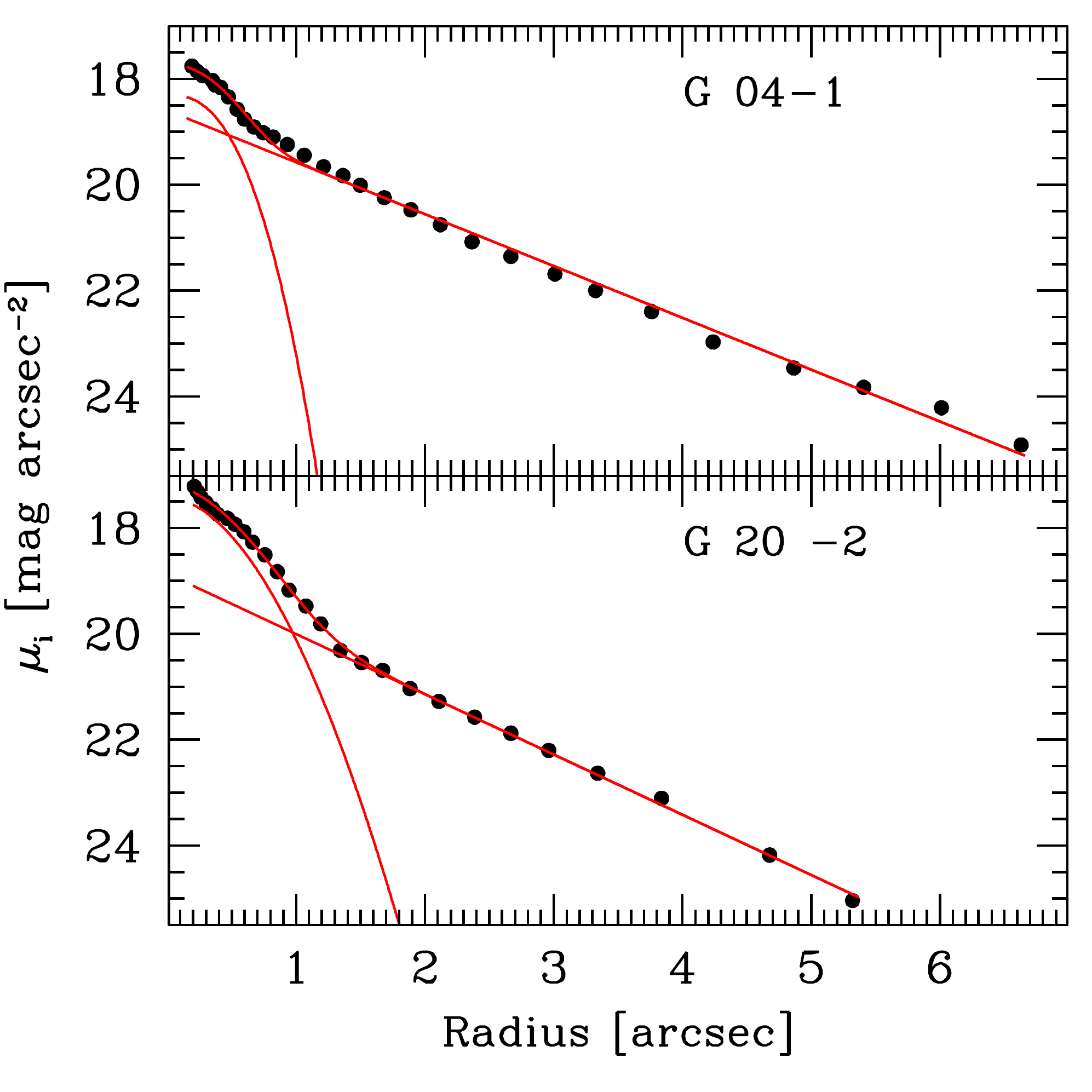}
\caption{Surface brightness profiles of continuum emission for the two
  target galaxies, G~04-2 (top) and G~20-2 (bottom). The flux of the
  continuum maps of both profiles has been scaled to match that of
  SDSS~$i$ band profile.  The solid dots
  represent isophotes and the red line represent a S\'ersic function bulge plus
  exponential disk fit to the surface brightness profile. For each
  galaxy there is a ``bulge'' in the surface brightness profile that
  is co-located with the rise in velocity dispersion in our kinematic
  maps (Figure~\ref{figure:gaskine}). Some galaxy properties from these
  fits are included in Table \ref{table:physparam}. \label{figure:expprof}} 

\end{figure}

In Figure~\ref{figure:expprof}, we show the surface-brightness profiles
of the star light for both target galaxies. These profiles are composite
surface-brightness profiles of HST FR647M, SDSS $i$ and continuum
emission from our GMOS observations (see
Figure~\ref{figure:clumps}). The surface photometry from each
telescope is scaled and combined using the method of \cite{fisher2008}. This method reduces random
errors (eg.,~sky subtraction) and allows us to probe
surface-brightness profiles to larger radius. The stellar emission is
smooth, and the main body of both 
galaxies ($r\ge 1.5$~kpc) are well described by exponential
surface-brightness profiles at large radii, suggesting that these are unperturbed
disks \citep{vdkAA11}. We determine scale lengths of 2.75 and 2.63 kpc
for G 20-2 and G 4-1 respectively, these values are also included in
Table \ref{table:physparam}.
Both galaxies have bulges in their continuum
emission. We indicate our bulge-disk decomposition with red lines. The
bulges in these galaxies
appear to be co-located with rises in the velocity-dispersion maps of
Figure~\ref{figure:gaskine}. The central kinematic component will be
discussed in more detail in Section~\ref{section:globalsigma}.

\begin{figure}
\includegraphics[width=\linewidth]{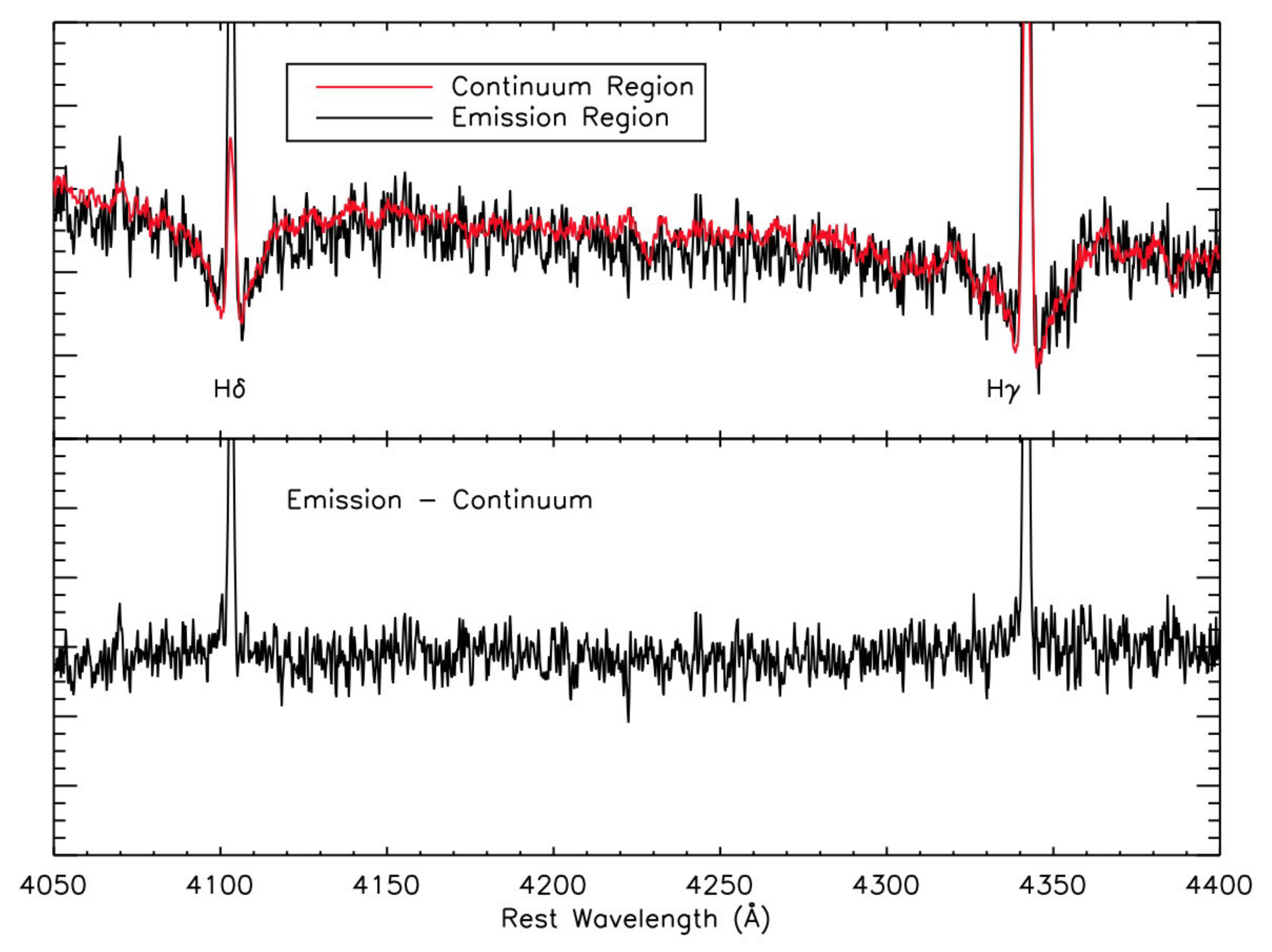}
\vspace{0pt}
 \caption{Comparison between spectra observed in the continuum peak
   and the brightest H$\beta$ clump of G 20-2, identified from our GMOS
   datacubes, summed in a $0\farcs5$
   aperture centered
   on these regions and normalized by the median continuum
     level. Our spectra are characterized by strong emission lines, in
     particular [OIII] and Balmer lines, as well as strong Balmer
     absorption associated with A-type stars. The top panel shows the continuum peak spectrum in red
   over the emission line peak spectrum in black, while the bottom shows the
   continuum peak spectrum subtracted from that of the emission line
   peak.  Here we show a section of our observed spectral
   range which includes the H$\gamma$ and H$\delta$ absorption lines,
   which (along with H$\beta$) are the strongest we observe. The shape of the Balmer absorption
   lines in both are quite similar, suggesting that the underlying
   stellar populations in both regions are of comparable age. The major
   difference between the spectra is the equivalent width of the
   emission lines as seen in the bottom panel. This is
   not surprising as the regions in which the emission lines are
   strongest are not coincident with the peak in continuum flux.} \label{figure:cspecs}
\end{figure}

In Figure \ref{figure:cspecs}, we make a qualitative comparison between
the spectra in different regions. We show two spectra, each
in artificial apertures with diameter equal to $0\farcs5$. The first is centered on the continuum peak and
the second is located at the peak of H$\beta$ flux from our GMOS datacubes. Figure \ref{figure:cspecs} shows a small
section of these spectra centered between H$\delta$ and
H$\gamma$ for galaxy G 20-2 (spectra for G 4-1 are qualitatively very
similar). These spectra have been normalized by the median continuum
value for direct comparison. Although not apparent in Figure
\ref{figure:clumps}, Figure \ref{figure:cspecs} clearly shows that we
observe significant line emission even in the central regions. The one
difference between the two regions is a much higher emission-line
equivalent width in the emission-line region, apparent in the
difference plot (bottom panel of Figure \ref{figure:cspecs}). This is not surprising
as the regions in which the emission lines are strongest are not
coincident with the peak in continuum flux. 

The emission-line-region spectrum has a lower S/N ratio due to
the lower continuum surface brightness, but the overall
continuum shape is nearly identical to that of the peak continuum
region. This suggests that the regions in which we can reliably
observe absorption-line profiles host similar aged stellar
populations. This could partially be an
effect of the seeing in which age differences are smeared
out. We also note that it is quite difficult to
probe stellar populations in the regions of maximum H$\beta$
equivalent width (where the average stellar age is likely to be the
lowest) due to the exponentially-dropping continuum surface brightness.

By examining the ionized gas and continuum components of G 4-1 and G
20-2 we provide further evidence that these galaxies are excellent
low-redshift analogs to high-redshift clumpy disks. We find that the
ionized gas is arranged in multiple clumps surrounding a central
continuum peak. We show that clumps are marginally
resolved, but that our 0$\farcs$1 HST imaging will be required to
measure accurate sizes for these clumps. Surface brightness profile
fitting to combined HST, GMOS, and SDSS data shows that both galaxies
are well described by an exponential at large radii with evidence of a
small bulge.\\

\begin{figure}
\includegraphics[width=\linewidth]{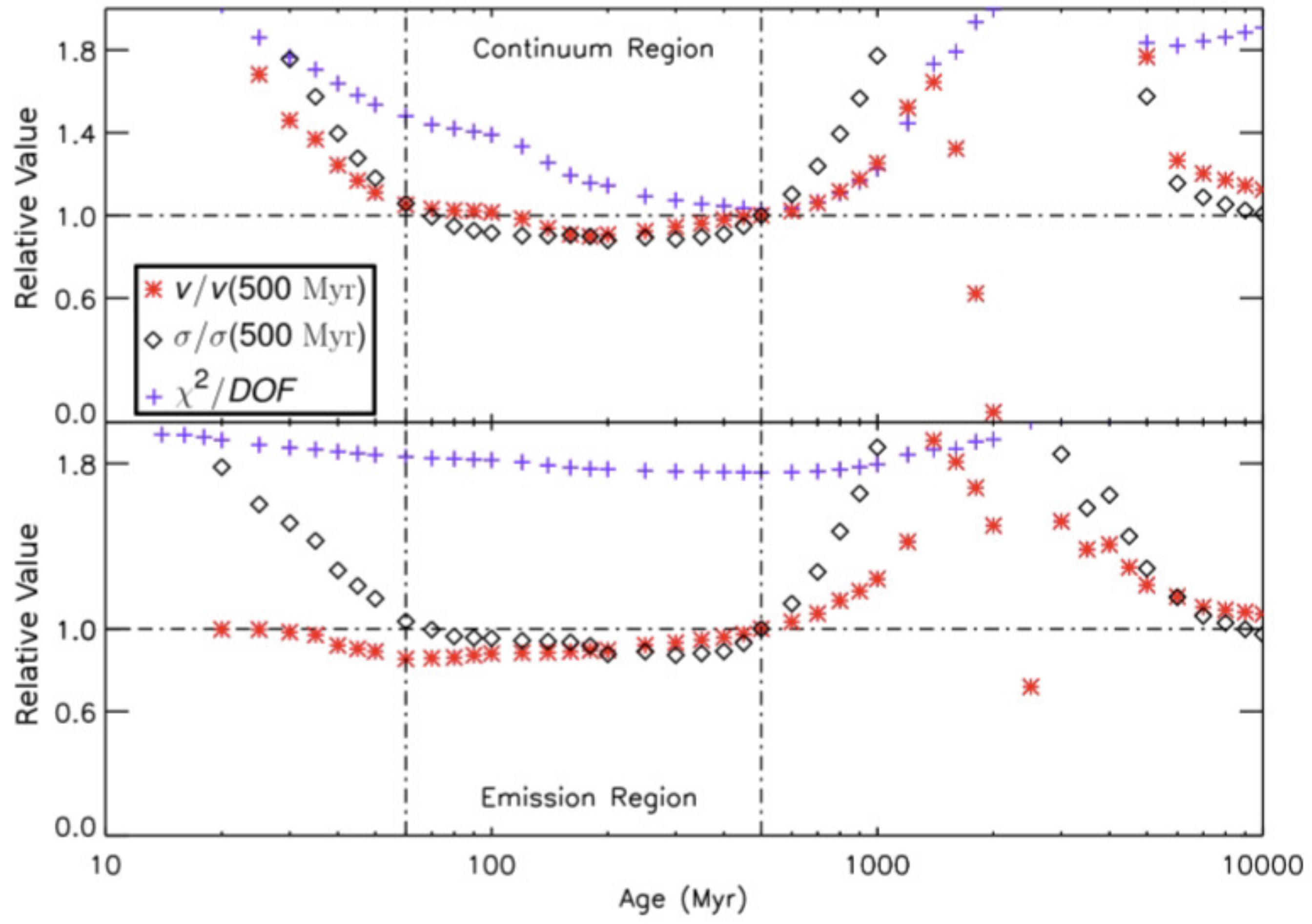}
 \caption{Velocity and velocity dispersion measurements of stellar
   kinematics for spectra shown in Figure \ref{figure:cspecs} at
   each template age (as described in Section
   \ref{section:tempage}). Both kinematics measures are
   normalized by measurements
   made using templates with ages of 500 Myr. We have overplotted a dot-dashed
   line on the vertical axis at 1 for reference. We find in the age
   range from 60 to 500 Myr (indicated by vertical dot-dashed lines)
   measurements of stellar kinematics remain relatively stable. We
   have also overplotted the $\chi^{2}$/DOF (not normalized) for each
   template and find 
   that this range also corresponds to the statistically best fitting
   templates. We adopt a template age of 500 Myr for fitting stellar
   kinematics in this study as this corresponds to the $\chi^{2}$/DOF
   closest to 1 for both regions.} \label{figure:agetst}
\end{figure}

\subsection{Matching Templates with Observed Spectra}\label{section:tempage}

We employed the full spectrum
stellar-kinematics fitting program Penalized Pixel-Fitting
\citep[pPXF,][]{cappe04} to measure the velocity and velocity
dispersion of the 
stellar populations in our sample. We fit our observations using synthetic galaxy spectra created with the
P\'{E}GASE-HR code \citep{leborg04}. This code uses high-resolution
(R=10,000) stellar spectra from the \'{E}LODIE library \citep{prug01} to
construct an evolutionary sequence of composite spectra based on simple stellar
populations (SSPs) with ages ranging from 10 Myr to
20 Gyr. 

We first developed a test to select the P\'{E}GASE template with an SSP age
most similar to our observed spectra.
Selecting a spectral template that is well matched to the observed
spectra is essential in reliably extracting kinematics through full
spectrum fitting. This is because of the wide range of absorption line profiles
associated with different stellar populations. Our test involves Using
pPXF to fit the 
representative spectra considered in Section 3.1 (see Figure
\ref{figure:cspecs}) with each P\'{E}GASE-HR template individually while
masking the emission lines. This procedure measures the
velocity, velocity dispersion, and goodness of fit ($\chi^{2}$/DOF) at
each SSP template age. We performed this test on spectra from both
the continuum and emission line peaks to check for any spatial
variation in stellar population age. We then selected the template with
$\chi^{2}$/DOF closest to 1. The results of template selection are
shown in Figure \ref{figure:agetst}. Kinematic values presented have been
normalized by the measurement made using a template age of 500 Myr. 

Figure \ref{figure:agetst} shows that there is a range in
template ages between 60 and 500 Myrs in which both the stellar
velocity and velocity dispersion measurements remain relatively
stable. The $\chi^{2}$/DOF values indicate that templates in this age range are
the best fitting from the P\'{E}GASE-HR library (this is also
apparent from visual inspection of individual fits). This is true for
both the continuum and emission line peaks suggesting that there is no
evidence of spatial variations in stellar population ages in our
observations. 

We investigated the effects of repeating our stellar kinematics
fitting (results presented in Figure \ref{figure:sk202}) using each
templates in the age range from 60-500 Myr individually. 
Selecting templates in the age range of 100-500 Myr returns velocity and
velocity dispersion measurements in agreement with measurements using
a 500 Myr old template within our simulated uncertainties (discussed
in Section \ref{section:uncert}). At younger template ages, the
velocity remains in agreement while the velocity dispersion
measurement increases relative to measurments made using 500 Myr old
templates. This difference increases as template age decreases from a
$\sim$5\% increase in velocity dispersion using a 100 Myr template to
a $\sim$38\% increase from fits performed with a 60 Myr template.

Qualitatively it
is reasonable to expect the continuum emission from these galaxies in
this wavelength regime to
be dominated by young stellar populations because the extreme levels
of line emission are indicative of a recent (and likely ongoing) star formation
activity. A P\'{E}GASE-HR template with an age of 500 Myr is selected
for stellar kinematics measurements throughout the remainder of
this paper. This was chosen as fits using templates aged 500 Myr
display the nearest $\chi^{2}$/DOF value to 1 in both the continuum
and emission line regions for both galaxies. 

\subsection{Measuring Kinematics}\label{section:measuringkinematics}

\subsubsection{High Resolution Ionized-Gas Kinematics}\label{section:gkin}

\begin{figure}
\includegraphics[width=\linewidth]{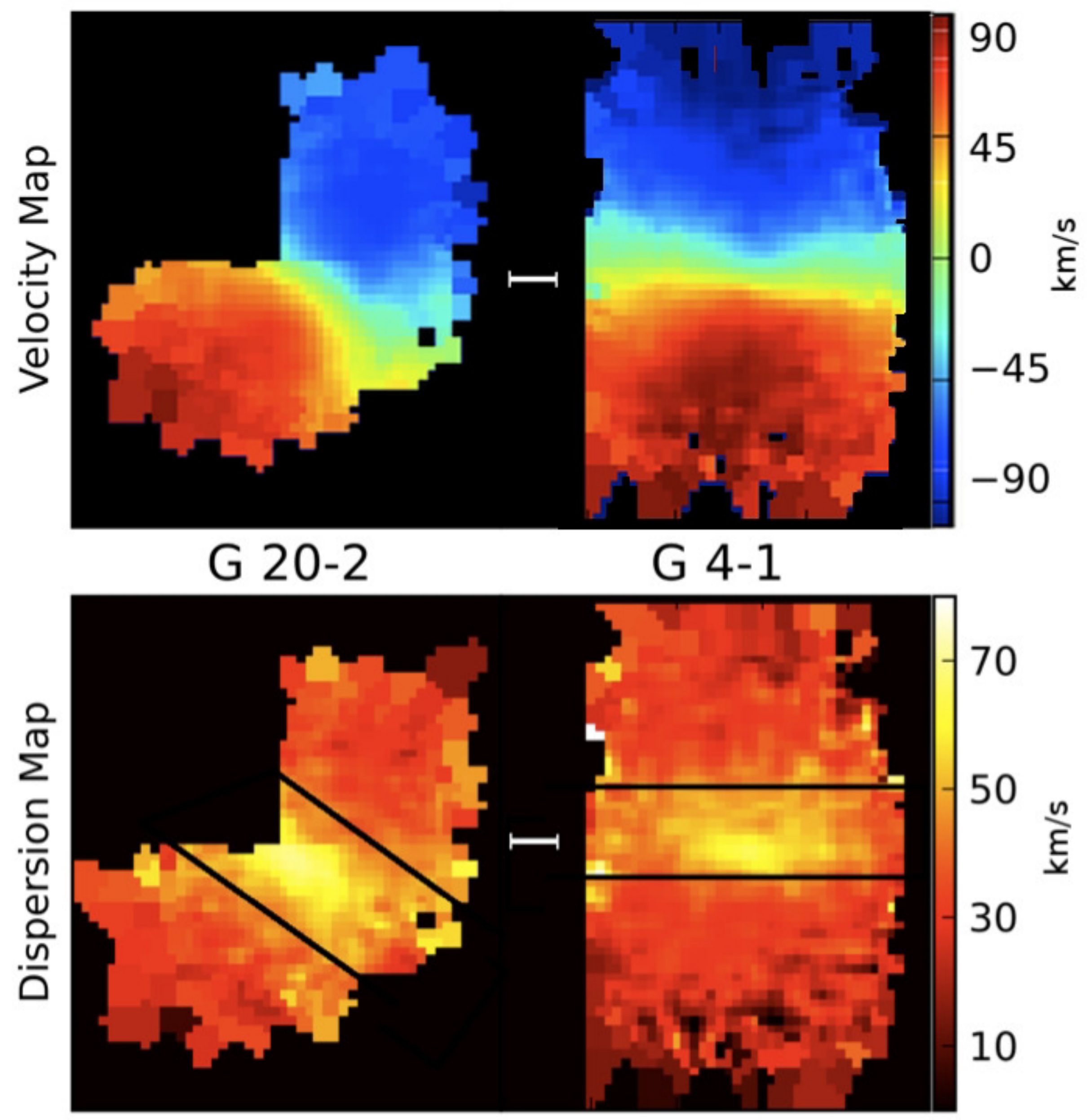}
\caption{Velocity and velocity dispersion maps for the ionized gas
   component of our sample measured by fitting a Gaussian profile to
   the H$\beta$ emission line. The small white bars
   represent a distance of $\sim1.5$ kpc at each galaxy's redshift. 
   Both galaxies display smoothly varying velocity fields and centrally
   peaked velocity dispersions typical of disk galaxies. The velocity
   dispersions here are large relative to typical disk galaxies
   nearby \citep[5-20 km s$^{-1}$ from DISKMASS, ][]{andersen06}  suggesting that the gas disks in these
   galaxies are highly turbulent. The black lines
   shown in the bottom row mark the boundaries of the ``central
   region'' which exhibits increased velocity dispersion. These
   regions possibly represent a young galactic bulge while velocity
   dispersions in the outer bins are characteristic of a
   marginally stable, turbulent disk. We measure a mean ionized gas
   velocity dispersion of 36.6 km s$^{-1}$ and 29.8 km s$^{-1}$ for
   the outer regions of G 20-2 and G 4-1 respectively. See Section
   \ref{section:globalsigma} for further discussion of
   global $\sigma$ estimates.}\label{figure:gaskine}

\end{figure}

We produced ionized gas velocity and
velocity dispersion maps by fitting a single
Gaussian profile to the H$\beta$ emission line. The velocity and
velocity dispersion are then calculated using the centroid position
and $\sigma_{gas}$ of the model fit. To correct for instrumental broadening,
we subtracted the instrumental line width from the raw measurement in
quadrature: $\sigma_{gas}=\sqrt{\sigma_{obs}^{2}-\sigma_{inst}^{2}}$,
where $\sigma_{obs}$ is the observed velocity dispersion and
$\sigma_{inst}$ his the instrumental broadening measured from our arc
observations to be 24 km s$^{-1}$. Figure \ref{figure:gaskine} shows
the results of this procedure. 

We employ the Voronoi Binning method
of \citet{cappe03} in
order to extend this analysis to radii beyond which these measurements
from single spaxels are reliable. This technique uses maps of S/N to
ensure that each spatial bin falls above a specified S/N
threshold. For our inputs we use the continuum maps of our galaxies
from Figure \ref{figure:clumps} and a constant noise based on the
variance of our spectra in off target regions. Because the line
emission is significantly stronger than that of the continuum we use a
low S/N threshold of 1.0, however many bins (particularly
those at the center) have S/N much higher than this. We choose to use
continuum S/N rather than emission-line S/N because the clumpy
structure results in oddly shaped bins which distort the kinematic
maps. Our low S/N threshold ensures that within the region in which we
observe clumps, all Voronoi bins contain a single spaxel.

\begin{figure}
\includegraphics[width=\linewidth]{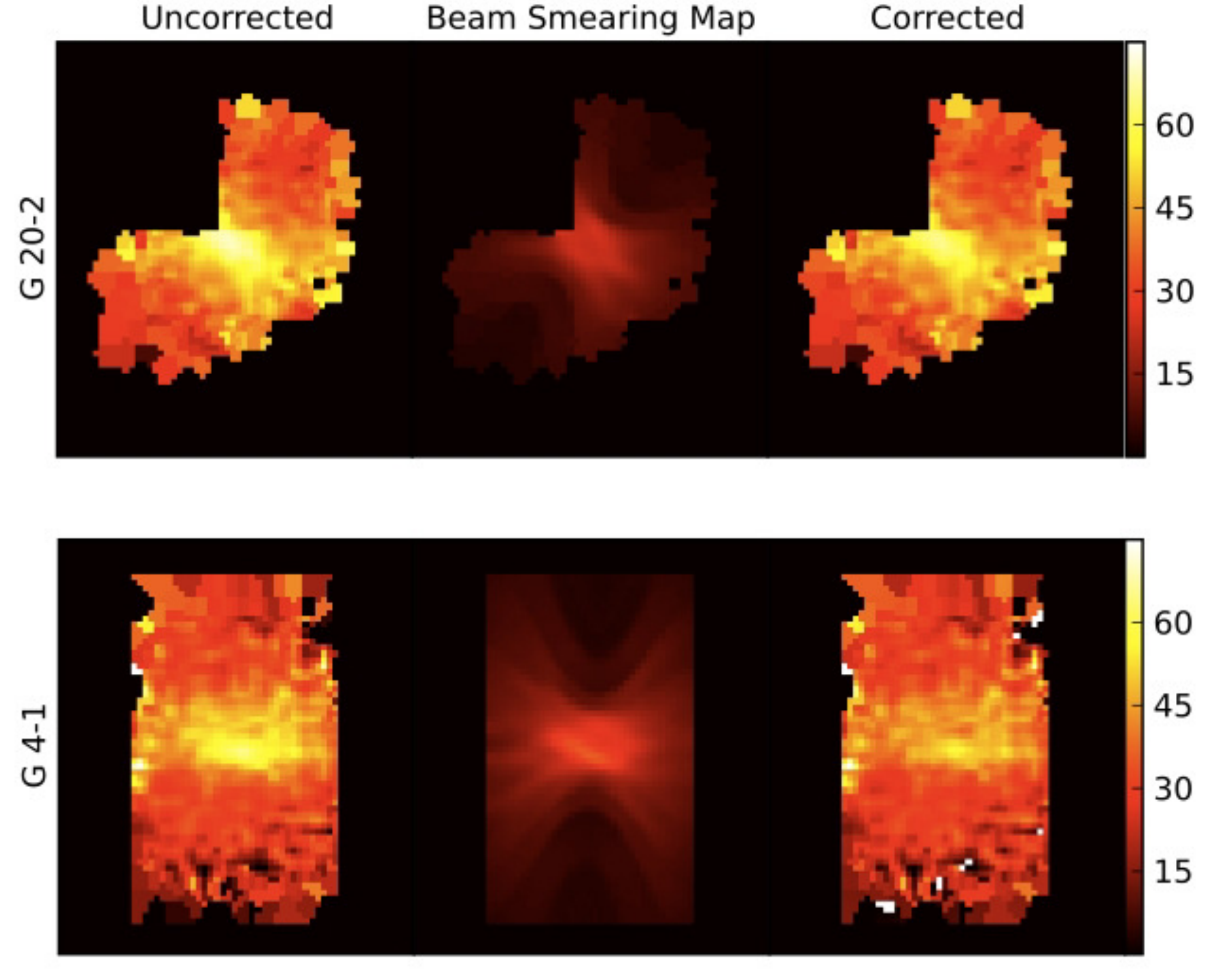}

\vspace{0pt}
 \caption{Modeling the effects of beam smearing. Plots for each disk
   galaxy are shown in a single row. From left to right each row shows
   maps of velocity dispersion corrected for instrumental broadening,
   a simulated map of the influence of beam smearing (described in
   Section \ref{section:beamsmearing}), and velocity dispersion maps
   corrected for beam smearing by subtracting the central panel from
   the left panel in quadrature. It is seen that the
   influence of beam smearing peaks at the centers of our galaxies as
   expected. Beam smearing estimates in these regions are 24-36 km
   s$^{-1}$ and the peak measured velocity dispersions are $\sim70$ km
   s$^{-1}$. The applied correction here is around 4-8 km
   s$^{-1}$ showing that even at the centers of our galaxies this is a
   minimal effect. In the outer portions of our disks the influence of beam
   smearing can effectively be ignored.}\label{figure:bscor}

\end{figure}

We used our kinematics maps to measure global $\sigma$ values which we
compare with previous results.
The ionized-gas-velocity maps we produced from our GMOS observations
extend into the flat portion of the rotation curve (which is shown
later in Section \ref{section:results}) where beam-smearing effects are
minimized. This allows for robust measurements of velocity
dispersion. We took an average value of the outer bins to make an
estimate of the velocity-dispersion value across the disk. We 
indicate the bins over which these averages were
taken by black lines in the bottom panel of Figure \ref{figure:gaskine}.
These bins correspond to regions that are a distance equivalent to our
seeing away from the kinematic centers of our galaxies. This results in mean H$\beta$ dispersions of
36.6 km s$^{-1}$ and 29.8 km s$^{-1}$ for G 20-2 and G 4-1,
respectively. We consider these our best estimates of the overall disk
velocity dispersion and compare them with velocity disperions of local
disks in Section \ref{section:gdiscuss}.

We also took a raw averages without excluding central bins to compare with previous work \citep{green13}. This results in values of 43.2 km
s$^{-1}$ and 35.6 km s$^{-1}$. For G 20-2 this is much closer to the
value of 44.9 km s$^{-1}$ found from SPIRAL data, while the
dispersion measured for G 4-1 is significantly lower
($\sigma_{SPIRAL,m}$ = 50.2 km s$^{-1}$). 

We performed a test to see if the difference between our GMOS and
SPIRAL results is an effect of spatial resolution.
We resample our GMOS datacubes to the resolution of the previous
SPIRAL observations and follow the same Gaussian fitting procedure
outlined above. We then calculated a flux weighted velocity dispersion
(matching the analysis of \citealp{green13}) for both galaxies
resulting in $\sigma_{m}$ of 45.3 and 38.8 km s$^{-1}$ for G 20-2 and
G 4-1. For G 20-2, $\sigma_{m}$ is well matched to the value calculated from
SPIRAL data, while our measurement for G 4-1 is still 10 km s$^{-1}$
less than previous measurements. 

The remaining difference of 10 km s$^{-1}$ between GMOS and SPIRAL
measurements for G 4-1 (as well as the
10-20~km~s$^{-1}$ difference between $\sigma_{m}$ of \citet{green13}
and our mean $\sigma_{gas}$ excluding central bins) can be attributed to a bias of
the previous SPIRAL measurements towards regions of
highest H$\alpha$ flux due to the lower sensitivity of previous
observations. This means that our $\sigma_{m}$ values from SPIRAL observations contain
little or no contribution from dynamically cooler gas at large
radii. Additionally, if the most intense line emission is coincident with the kinematic center
of a galaxy, then our previous measurements will be more strongly
influenced by beam smearing as the central region is where velocity gradients will
be at a maximum. This is the case for galaxy G 4-1 while in G 20-2
the most intense line emission is offset from the kinematic center (see Figures
\ref{figure:clumps} and \ref{figure:gaskine}). 

In summary,
discrepancies between $\sigma_{m}$ measurements from SPIRAL and GMOS
can largely be attributed to spatial resolution, sensitivity, and beam
smearing. We note that our GMOS dispersions are consistent with higher resolution
kinematics maps based on Paschen $\alpha$ from adaptive optics
assisted observations using the OH-Suppressing Infrared IFS
(OSIRIS) on Keck (Bassett et al. in prep). A more detailed analysis of global velocity
dispersions can be found in Section \ref{section:globalsigma}.

\subsubsection{Modeling the Effects of Beam Smearing}\label{section:beamsmearing}

While the GMOS observations
presented here benefit from improved spatial resolution, beam smearing
can still play an important role in understanding our kinematic
measurements. \citet{davies11} shows how typical observations of
$z\sim0.1$ galaxies at $\sim$1$\farcs$0 resolution can inflate the
velocity dispersion of average disk galaxies \citep[$\sim10-25$ km
s$^{-1}$,][]{andersen06,martin13} to the large values found here.

Beam smearing affects both the observed velocity and velocity-dispersion
maps due to velocity gradients over individual beams. The end
result is a slight reduction in the slope of rotation curves and an
inflation of velocity-dispersion measurements. This effect is most
pronounced in the central regions of disk galaxies, where velocity
gradients are at a maximum. We can attempt to correct for this by
carefully considering the observed shape of the velocity field as well
as the average seeing of our observations.

We performed a simulation to better understand quantitatively the
effects of beam smearing on our velocity-dispersion measurements.
We first fit a rotating
disk model to the observed ionized gas kinematics, the details of which
can be found in \citet{green13}. We mapped the influence of
beam smearing on these models when subject to conditions
typical of our observations ($\sim0\farcs5$ seeing). This mapping was done by
creating an artificial datacube with two spatial dimensions
and one velocity, with the velocity resolution chosen to
match our observations. The velocity distribution of each
artificial spaxel received a contribution from every other spaxel 
weighted by a two dimensional Gaussian function with $\sigma$ equal to
our average seeing. The velocity
distribution was subsequently fit with a Gaussian function and the
$\sigma$ of these fits were taken as the contribution of beam smearing
to our measured velocity dispersions. 

We subtracted the produced ``beam smearing maps'' in quadrature from the
observed velocity-dispersion maps to correct for the beam-smearing
effect in our velocity dispersion maps \citep{gneru11,epinat12}. Maps of our modeled beam smearing
and the corrected velocity dispersion are shown in
Figure \ref{figure:bscor}. From the central figure in each row it is
seen that the estimated influence of beam smearing peaks as expected
in the central region, where the velocity gradient is at a
maximum. Typical values in the central regions of our beam-smearing maps are 24-30 km s$^{-1}$, which is low
relative to measured velocity dispersions in the same regions (55-80
km s$^{-1}$). 

\begin{figure*}
\includegraphics[width=\linewidth]{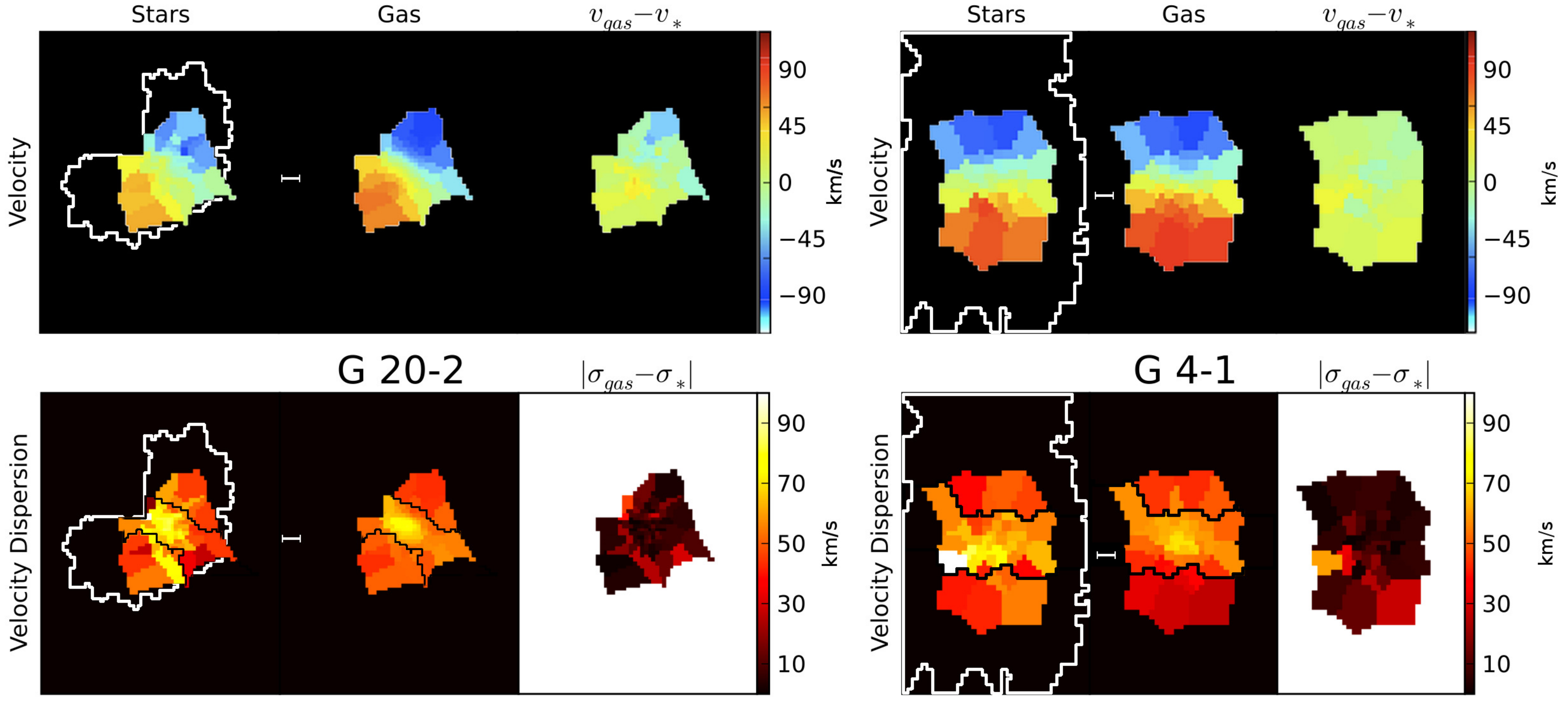}
\vspace{0pt}
 \caption{Results of pPXF measurements of stellar kinematics in our
   sample. In order to obtain reliable measurements in the faint
   outskirts of these galaxies, the Voronoi Tessellated binning method
   was employed. The gas velocity was remeasured in each bin for
   direct comparison. White bars in each panel correspond to a
   distance of $\sim1.5$ kpc at the redshift of our
   galaxies. The velocity maps are plotted over the footprint of the
   gas only measurements for comparison. The gas and stars in these
   galaxies appear to be closely kinematically coupled, both are arranged in
   a smoothly rotating disk with a large velocity dispersion. We
   calculate a mean stellar velocity dispersion value for both the gas and
   stars, excluding central regions (indicated by black lines in the
   velocity dispersion maps) for use in simple disk models. This
   results in values of
   $\sigma_{d,*}=52.8$ for G 20-2 and $\sigma_{d,*}=47.5$ for G
   4-1. See Section \ref{section:globalsigma} for further discussion
   of global sigma values} \label{figure:sk202} 
\end{figure*}

We also compared the central beam-smearing estimates with a simple analytic
calculation. We assumed a linear velocity gradient and constant
intensity, which is a reasonable approximation for the central regions
of smoothly rotating disks. We estimated a central velocity gradient
from Figure \ref{figure:gaskine} of 5-6 km s$^{-1}$ pixel$^{-1}$ by
measuring the difference in velocity between the two flat portions of
the rotation curves and dividing by the separation between these two
portions in pixels. The contribution of beam smearing was then estimated as

\begin{equation}\label{bsanal}
  \sigma_{BS} \approx \frac{dV}{d\theta}\sigma_{\theta},
\end{equation}

where $d\theta$ is a distance corresponding to one pixel and $\sigma_{\theta}$ is the
seeing of our observations in pixels ($\sim$5 pixels). This results in
a value of 25-30 km s$^{-1}$, consistent with the modeled values. 

When the modeled beam-smearing contribution is subtracted in
quadrature from the observed velocity dispersion in 
the regions of maximum velocity dispersion (55-80 km s$^{-1}$), the
correction is only 4-8 km s$^{-1}$, comparable to our errors in
stellar-kinematics measurements (see Section \ref{section:uncert}). This
small correction suggests that the increased velocity dispersion in the
central regions are real and a possible signature of a small bulge or
pseudobulge (see Section \ref{section:clumpfate} for discussion). 
Furthermore, in regions further than our seeing distance from
the central regions, beam smearing can effectively be ignored. 

\subsubsection{Mapping Stellar Kinematics}\label{section:skin}

In this section, we map the stellar kinematics of our galaxies and
compare them directly to those of the ionized gas. We again employed the Voronoi Binning method
of \citet{cappe03} to ensure that our measurements are reliable even in the faint
outskirts of our galaxies. We used the same setup as in
Section \ref{section:gkin}, but with a larger S/N
threshold of 10. We used a larger S/N threshold here because the absorption signal
is much weaker than the emission line strength.

We developed a custom IDL wrapper for pPXF which simultaneously fits for
stellar and emission-line kinematics. This allows us to directly compare stellar and gas
kinematics on a bin-by-bin basis. We first fitted the spectrum with pPXF using
emission-line masking and subtracted the resulting fit.  The result
is a spectrum composed only of emission lines. Next, we 
fit the emission lines using a single Gaussian profile. From
the centroid position and $\sigma$ of these fits we again measured the
velocity and velocity dispersion of the ionized gas in our galaxies.

We plot the results of our stellar-kinematics fitting procedure in
Figure \ref{figure:sk202}. For each galaxy, the velocity
maps of the ionized gas and stars, as well as the subtracted
residual of $v_{gas}-v_{*}$, are shown in the top row. Similarly,
the bottom row shows the maps of velocity dispersion, but this time
showing the absolute difference: $|\sigma_{gas}-\sigma_{*} |$. This
allows us to plot using the same color bar used for the
velocity-dispersion maps, however we note that in most bins the
stellar velocity disperions is larger than that of the gas (see Figure
\ref{figure:rotcurves}). Also
depicted in the top row of Figure \ref{figure:sk202} as solid white lines are the
footprints of the gas only kinematics measurements from Figure
\ref{figure:gaskine} for comparison. These footprints indicate the
extent to which we can reliably measure ionized gas kinematics. Solid
black lines in the bottom 
row differentiate bins that are considered part of the central region and
those that are part of the outer disk. Global $\sigma$ values are
calculated for these regions and discussed in Section
\ref{section:globalsigma}.

\subsubsection{Stellar Kinematics Uncertainties}\label{section:uncert}

We next consider if the residual-kinematic maps (right columns for
both galaxies in Figure \ref{figure:sk202}) are significant or if they
are within our measurement uncertainties.
Following the procedure suggested by
\citet{cappe04}, we performed a Monte Carlo simulation of our absorption fitting
routine. This simulation was done by first subtracting the best-fit spectrum
output by pPXF, from the original spectrum, to produce a 
residual spectrum. The residual spectrum was then randomized in wavelength (using the IDL task SHUFFLE), and added to the best
fit in order to simulate an idealized spectrum with noise
characteristic of our observations. We chose to add the shuffled
residual to the model rather than the original spectrum as the latter
tends to produce spectra which are noisier than the original, thus
overestimating the uncertainty. Some areas of the
spectra display unusually large residuals due to the chip
gaps in the detector (which are masked out in the fitting procedure).
To account for these large deviations we replace values in
masked regions by random values based on the standard deviation of the
unmasked regions of the residual spectrum.

\begin{figure}
\includegraphics[width=\linewidth]{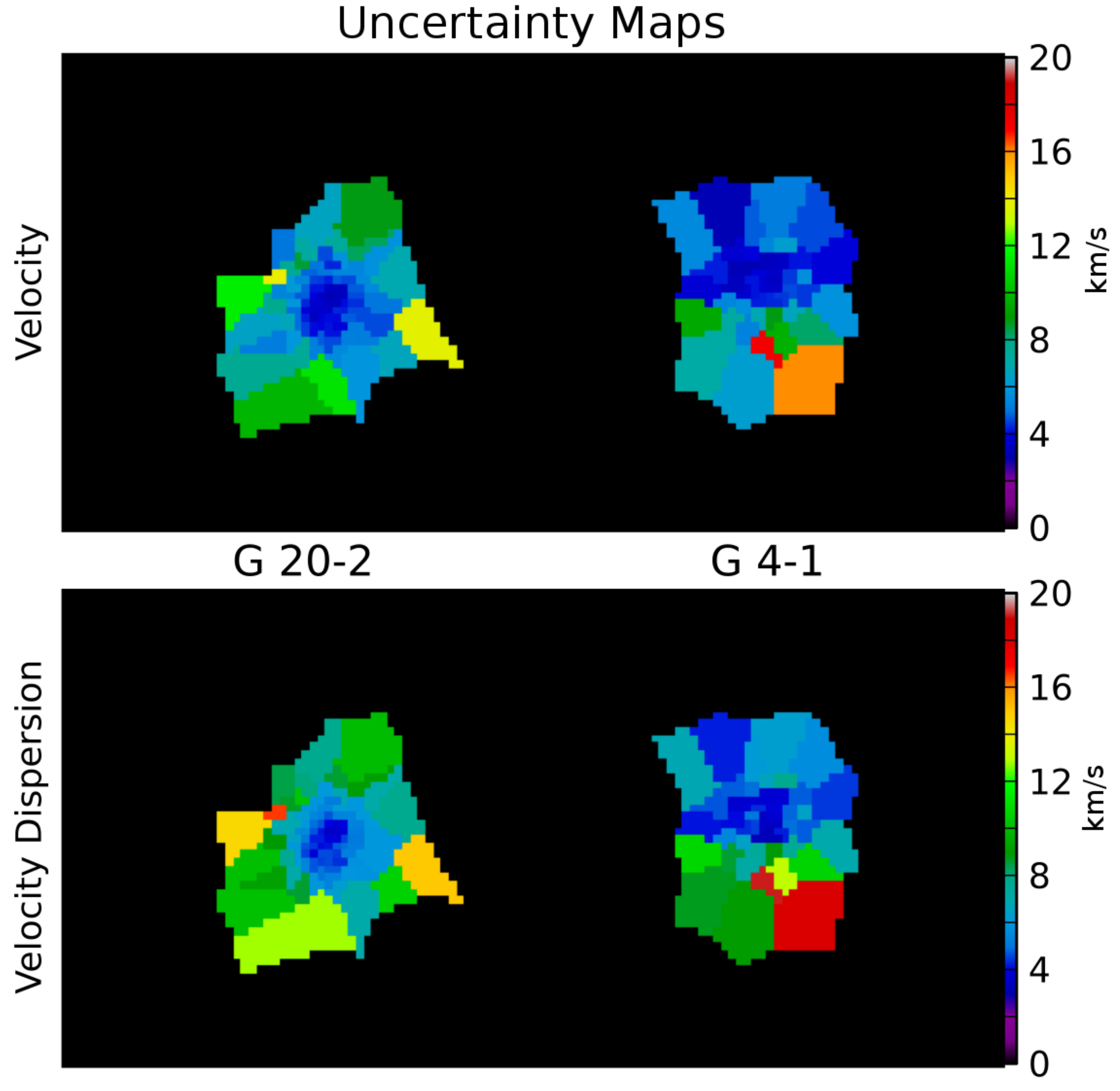}

\vspace{0pt}

 \caption{To estimate the uncertainties of our pPXF measurements, we perform
   a Monte Carlo simulation as suggested by \citet{cappe04}. Briefly
   the residual of our stellar kinematics fits are shuffled and added
   to the best fit model. We then refit this spectrum with the
   original model in an attempt to recover the known input
   kinematics. This process is repeated 1000 times for each spatial
   bin and the uncertainties are taken as the standard deviation of
   these measurements. Due to the high resolution of our spectra and
   the sensitivity of our observations, we are able to make
   measurements of stellar kinematics accurate to $5-8$ km s$^{-1}$ in
  most bins.} \label{figure:uncmaps}

\end{figure}

\begin{table}
  \caption{Stellar Kinematics Uncertainties}
  \centering
  \begin{tabular}{ l c c }
  \hline
   & Velocity Uncertainty & $\sigma$ Uncertainty \\ 
   & $\pm$ km s$^{-1}$ & $\pm$ km s$^{-1}$ \\
  \hline
  G 4-1 center & 4.7 & 5.2\\
  G4-1 disk & 6.7 & 7.9\\
  G 20-2 center & 5.0 & 5.9\\
  G 20-2 disk & 6.7 & 8.2 \\
  \hline \\
  \end{tabular}\label{table:uncert}
\end{table}

We produced 1000 simulated spectra
for each spatial bin and refit each of them using the same pPXF
setup. We then computed the standard deviations of the sample of
simulated fit values (using the square root of the IDL VARIANCE task).
These values were taken as our uncertainties in each bin.
Figure \ref{figure:uncmaps} shows the maps
of our uncertainties in velocity and velocity dispersion for G 20-2
and G 4-1. 
Our measurements are most accurate in the centers of our galaxies,
where the continuum strength is the highest. The centers are
also the areas in which the effects of beam smearing are the
greatest. 

Due to the sensitivity and high spectral resolution of our observations,
even measurements of the stellar kinematics made at the largest radii
are quite robust. 
We estimate the measurement uncertainty of our pPXF procedure by
taking two mean values for both stellar velocity and velocity
dispersion. One measurement in the central regions and one in the
outer bins. Central and outer regions are defined as in Section
\ref{section:skin} and Figure \ref{figure:sk202}. These values are presented in Table
\ref{table:uncert}. Kinematics measurements in most bins are known within
$\pm8$ km s$^{-1}$ based on our Monte Carlo simulation.

Comparing
Figures \ref{figure:sk202} and \ref{figure:uncmaps} one can see that
kinematics for these two components match quite closely. The mean
values of $|v_{gas}-v_{*}|$ and $|\sigma_{gas}-\sigma_{*} |$ are 15.8
and 8.4 km s$^{-1}$ for G 20-2 and 8.9 and 8.3 km s$^{-1}$ for G 4-1
mostly consistent with our estimated stellar kinematics
uncertainties. The large velocity difference in G 20-2 between stars
and gas may be attributed to asymmetric drift. This is a phenomenon
observed in many late-type galaxies in which the stellar disk lags
behind the gas disk due to extra support from random motions
\citep[][some evidence for this can be seen in
Figure~\ref{figure:rotcurves}]{golu13}.

We
find that the ionized gas and stellar kinematics of these galaxies are closely
coupled and characterized by smooth rotation and large,
centrally peaked velocity dispersions. It is interesting that these
two components are so well matched kinematically while their morphologies
(shown in Figure \ref{figure:clumps}) appear very different.

\subsection{Global Velocity Dispersion
  Estimates}\label{section:globalsigma}

\begin{table}
  \caption{Global Beam-Smearing Corrected $\sigma$ Values}
  \centering
  \begin{tabular}{ c c c c c c c c }
  \hline\hline
  Galaxy & $\sigma_{m,\textrm{SPIRAL}}^{1}$ &
  $\sigma_{c,gas}^{2}$ & 
  $\sigma_{c,*}^{3}$ &
  $\sigma_{d,gas}^{4}$ & 
  $\sigma_{d,*}^{5}$ & \\ 
   & km s$^{-1}$ & km s$^{-1}$ & km s$^{-1}$ & km s$^{-1}$ & km
   s$^{-1}$\\
  \hline
  G 4-1 & 50 & 44 & 58$\pm$5 & 30 & 48$\pm$8 \\
  G 20-2 & 45 & 49 & 66$\pm$6 & 37 & 53$\pm$8 \\
  \hline \\
  \end{tabular} \\
  \raggedright
  $^{1}$Flux-weighted mean
   ionized-gas velocity dispersion measured from H$\alpha$ emission
   lines \citep{green13}\\
  $^{2}$Unweighted-mean ionized-gas velocity
    dispersion of central regions measured from H$\beta$ emission observed with
    GMOS-IFS\\
  $^{3}$Unweighted-mean stellar velocity
    dispersion of central regions from pPXF fits to GMOS-IFS data\\
  $^{4}$Unweighted-mean ionized-gas velocity
    dispersion of disk regions measured from H$\beta$ emission observed with
    GMOS-IFS\\
  $^{5}$Unweighted-mean stellar velocity
    dispersion of disk regions from pPXF fits to GMOS-IFS data\\
  \label{table:centralkine}
\end{table}

We find that our disk galaxies exhibit enhanced
velocity dispersions in their central regions, which can not entirely
be attributed to the effects of beam smearing. Regions of enhanced
velocity dispersion correspond to a central excess in stellar light
compared with a single exponential profile
(see Figure~\ref{figure:expprof}). This
is possible evidence for a galactic bulge and thus we calculated two
average velocity disperions for both the stellar and ionized gas
components: one mean in the central regions ($\sigma_{c}$) and one in
the outer regions ($\sigma_{d}$). These two values correspond to a
``bulge'' and ``disk'' mean $\sigma$ respectively.  
Different regions for calculating mean $\sigma$ values are indicated in the velocity dispersion maps of Figures
\ref{figure:gaskine} and \ref{figure:sk202} by solid black lines. We
also calculated the average influence of beam smearing in the central
and outer regions. This was done by taking an average of pixels from
our beam-smearing maps (Figure \ref{figure:bscor}) in each of the
Voronoi Bins of Figures \ref{figure:gaskine} and
\ref{figure:sk202}. We then average these binned-beam-smearing values
in the same way as was done for the velocity-dispersion measurements.
A correction is then applied to the respective mean
velocity-dispersion values as $\left(\sigma_{mean}^2 - \sigma_{BS}^2\right)^\frac{1}{2}$.

We first consider the mean velocity-dispersion values of the central bins,
$\sigma_{c}$. The resulting values are shown in Table
\ref{table:centralkine}. The measurements made are the mean
ionized-gas velocity dispersion from gas-only measurements
($\sigma_{c,gas}$) and the mean stellar
velocity dispersion ($\sigma_{c,*}$). We also include for comparison
$\sigma_{m,\textrm{SPIRAL}}$ which is the flux weighted mean ionized
gas velocity dispersion from \citet{green13}. In both
galaxies, the values of $\sigma_{c,*}$ are larger than
$\sigma_{c,gas}$. The differences are 17.7 and 14.5 km s$^{-1}$ for G
20-2 and G 4-1, respectively. This can be attributed
to the fact that in our simple mean each Voronoi bin is given the same
weight. Because our data cubes are more coarsely
binned during our pPXF procedure, the outer regions (where velocity
dispersion are lowest) are given less weight per unit area
in $\sigma_{c,*}$ than in $\sigma_{c,gas}$. When Voronoi bins
are also weighted by bin size, this difference is reduced by
half. 

We next make similar measurements considering the outer disk regions
of $\sigma_{d,gas}$ and $\sigma_{d,*}$. These
results are shown in Table \ref{table:centralkine}. 
Similar to the central regions, our
measurements of $\sigma_{d,*}$ are larger than $\sigma_{d,gas}$ by
16.2 and 17.7 km s$^{-1}$ for G 20-2 and G 4-1, respectively.
The difference in this case is likely due to the reduced
spatial coverage of our pPXF kinematics (see Figure
\ref{figure:sk202}) which are restricted to the largest $\sigma$
regions of the galactic disks. 

\subsection{Rotation Curves}\label{section:rcurves}

\begin{figure*}
        \centering
        \begin{subfigure}
          \centering
          \includegraphics[width=.8\linewidth]{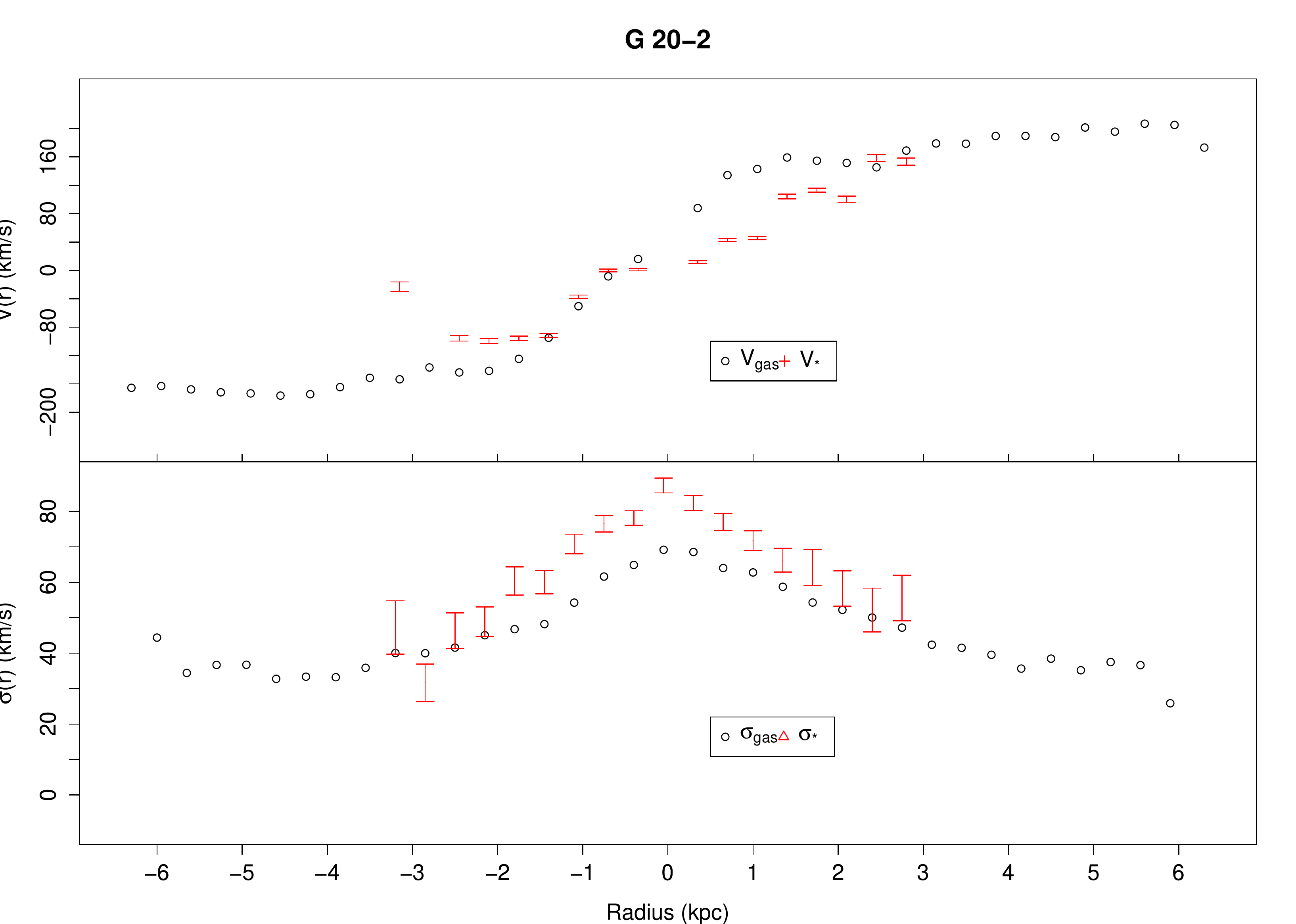}
        \end{subfigure}

        \begin{subfigure}
          \centering
          \includegraphics[width=.8\linewidth]{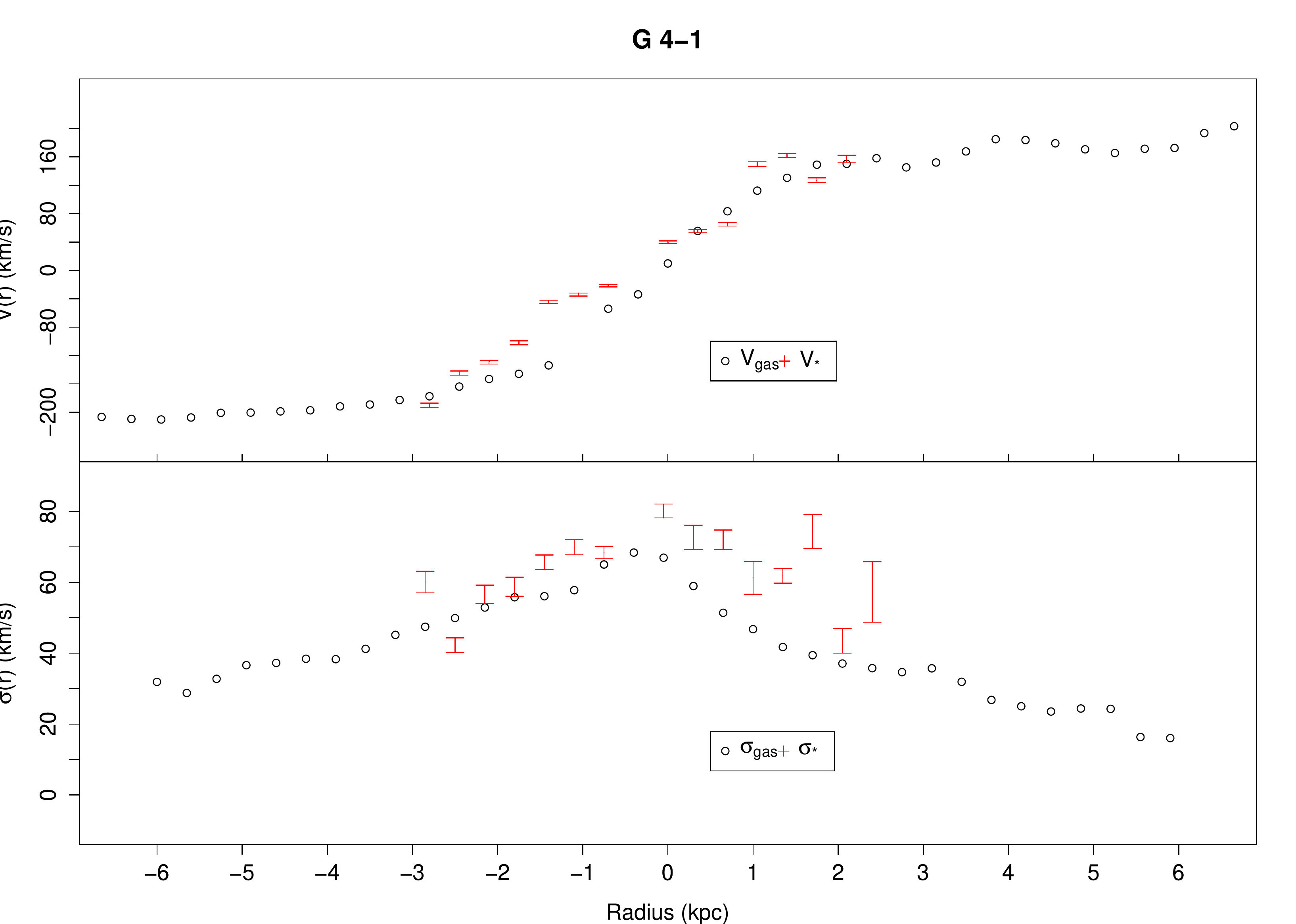}
        \end{subfigure}

        \caption{Galaxy rotation curves and $\sigma$
          profiles averaged in bins of 300 pc. Velocities are plotted
          with a correction for angular 
          dependence using $V(r,\phi)=V(r)\cos(\phi)\sin(i)$ where $\phi$ is the
          angle between a given spatial location and the semimajor axis with
          the center of the galaxy as the vertex and i is the galaxy
          inclination. For $\sigma$ we have 
          subtracted our beam smearing estimate from the observed
          velocity dispersion in quadrature on a bin by bin basis. We
          overplot values for the 
          emission line kinematics (black circles) and stellar
          kinematics (red error bars) extracted during our pPXF
          procedure. For the 
          emission line measurement, we probe well into the flat
          portion of the rotation curve where the effects of beam
          smearing are negligible while stellar kinematics
          measurements just reach this region. The size of the error
          bars plotted for our stellar kinematics indicate the average
          of our simulated uncertainties in each bin.}\label{figure:rotcurves}
\end{figure*}

The main result of this study is that in both galaxies,
\textit{the stellar kinematics closely mirror those of the ionized gas}. This
is shown again in Figure \ref{figure:rotcurves} where we plot our kinematics
measurements versus distance to the galactic center. In the top panels we
have computed $V(r)$ in each spatial bin using
$V(r,\phi)=V(r)\cos(\theta)sin(i)$ \citep{warner73} where $\theta$ is the azimuthal
angle in the galactic plane measured from the kinematic major axis and
$i$ is the inclination. In the bottom panels we plot velocity
dispersions after correcting for beam smearing by subtracting our
simulated beam smearing contribution (see
Section~\ref{section:beamsmearing}) from the measured velocity
dispersion in quadrature on a bin-by-bin basis. 

We estimated the inclinations of our galaxies, $i$, from axis ratios
determined using GALFIT \citep{peng02}. We performed this fit on
continuum images created from our data cubes but taking the median
value of each spaxel. The axis ratios calculated
using GALFIT are sensitive to the choice of point-spread function (PSF),
however there is little (if any) precedent for using GALFIT with IFS
data. We require a PSF which has been observed in a similar manner as
our galaxies. To achieve this we observed an A-type star with GMOS in
IFS mode with the spectral range chosen to match the rest frame
wavelengths of our galaxy observations. This data was reduced in the
same manner as described in Section \ref{section:dr}. We created a
median image of this star from the reduced data cube and used this as
our PSF while fitting with GALFIT. We consider this the best possible method
for defining the IFS response to a point source. We
report the axis ratios (b/a) we recover, as well as inclinations
calculated using the standard equation in Table \ref{table:physparam}. 

Figure \ref{figure:rotcurves} shows the computed rotation curves and
velocity dispersion profiles for our galaxies. Both the ionized-gas
(open circles) and stellar (red error bars) kinematics are measured
out to the flat portion of the rotation curves of these galaxies. This
allows us to make robust estimates of global velocity dispersions. In
all bins, the velocity curves of both components are well
matched with small differences likely due to differences in binning
between our gas and stellar kinematics fitting procedures. Large differences
seen in the rotation curves of G 20-2 are possible indications of
asymmetric drift. Stellar velocity dispersions in most bins agree with
the ionized gas velocity dispersions within our
simulated uncertainties (indicated by the size of the error bars) or
are found to be slightly higher. This suggests that the observed
turbulence in the ionized gas in DYNAMO galaxies is not primarily due
to star-formation feedback. If the turbulence was due to feedback,
stellar velocity dispersion measurements would be \textit{lower} than
that of the gas.

\subsection{Results Summary}\label{section:results} 

The optical spectra of these galaxies are dominated by young
stellar populations. Figure \ref{figure:cspecs} shows that
the absorption spectrum is dominated by Balmer lines. These lines are
strongest in
B and A type stars which dominate optical spectra of SSP's at ages
younger than 1 Gyr. This is consistent with the young ages (500 Myr)
of templates selected for our stellar kinematics fitting. These
young ages, along with the dynamically hot kinematics, high rates of
star formation \citep{green13}, and large gas fractions (Fisher et al.
submitted) are suggestive
of early stages of disk-galaxy evolution. Such an evolutionary state
is more characteristic of star-forming galaxies at high redshift than
star-forming galaxies at $z<1$. 

We map the relative strength of emission line and continuum flux
which reveals clumpy substructure which was not resolved in previous
observations (Figure \ref{figure:clumps}). Both galaxies exhibit
multiple regions of high emission-line 
equivalent width located 1-2 kpc from a single, centrally located peak
in continuum emission. The discovery of these
clumps lends further
support to the suggestion that these galaxies are low redshift
analogues to clumpy galaxies at $z\sim2$. 

We find that the kinematics of the stellar and ionized gas
components of these galaxies are well matched, with large stellar velocity dispersions
similar to ionized gas velocity dispersions found in previous work
\citep{green13}. This is indicative of a 
scenario in which the ionized gas turbulence is driven externally
rather than being primarily due to star-formation feedback. This
result is also found in a recent IFS study of galactic winds in local
luminous and ultraluminous infrared galaxies (LIRGS/ULIRGS)
\citep{arribas14}. Measuring winds from H$\beta$ using our GMOS
datacubes is inherently difficult due to the complex absorption
profile also associated with this transition. Emission line profiles
of [OIII] (5007 \AA) in galaxy G 20-2 show some evidence for broad
components which could be indicative of star-formation driven winds
expected in such high-SFR systems (see Section \ref{section:clumpfate}
for some discussion). Inclusion of a broad component
however, has a negligible effect on velocity dispersion measurements of
the narrow component when compared to a single component fit. 
Therefore not including a broad component in fits presented here will
not affect our conclusions. An
analysis of the broad components of emission lines of DYNAMO galaxies
will be presented in future work based on IFS
observations of Paschen $\alpha$ using Keck OSIRIS.

\section{Discussion}\label{section:discussion}

\subsection{Comparison With Local Disks}\label{section:gdiscuss}

\begin{figure}
\includegraphics[width=\linewidth]{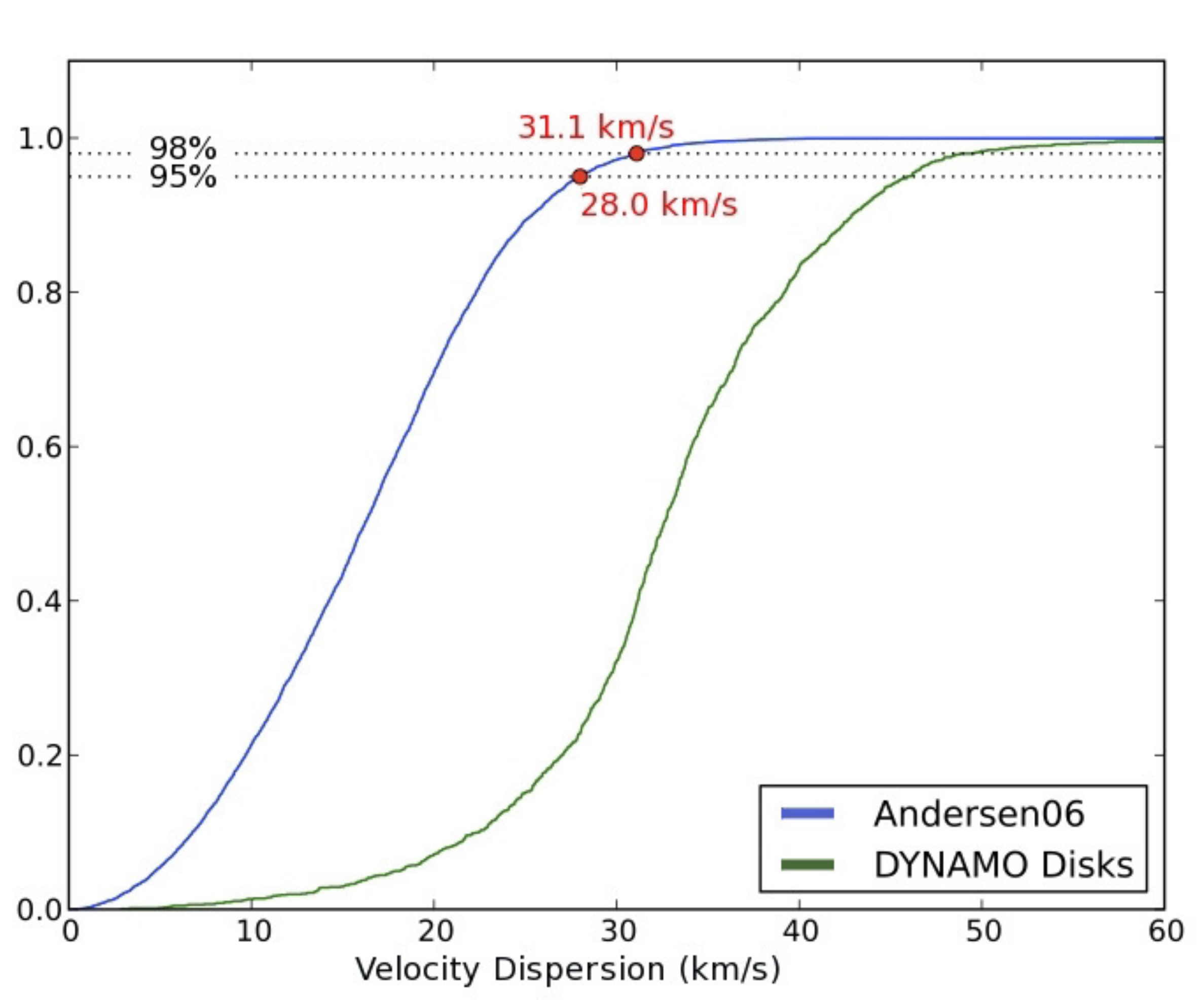}

 \caption{Cumulative fraction in $\sigma$ for each fiber from
   \citet{andersen06} compared with each value from outer spatial bins
   indicated in
   Figure \ref{figure:gaskine} for both galaxies considered here. It is clear that values measured for G
 20-2 and G 4-1 do not come from the same parent sample as those
 measured for local face on spirals. Indeed, the values of
 $\sigma_{d,gas}$ measured for these galaxies are in the $98^{th}$
 percentile of individual fibers from \citet{andersen06}.} \label{figure:loccomp}

\end{figure}

In Figure \ref{figure:loccomp}, we show that our ionized-gas velocity-dispersion measurements are
truly unusual compared to similar measurements of low-redshift late-type
galaxies. We established this difference based on comparison with the
IFS studies from the DiskMass survey of \citet{andersen06},
who examine the H$\alpha$ linewidths of 39 nearby, face-on disks using
the DensePak instrument on the WIYN 3.5 m telescope. We obtained
H$\alpha$-line-width measurements from each fiber from their observations
to compare directly with our H$\beta$ measurements in each of our
spatial bins from Figure \ref{figure:gaskine}. We plot in Figure
\ref{figure:loccomp} the cumulative fractions of velocity dispersion
for the two samples.

It is clear from Figure \ref{figure:loccomp} that the disk
galaxies considered here are outliers relative to local disk
samples. The disk velocity dispersions of $\sim32$ km s$^{-1}$ we measure for
our sample are in the $98^{th}$ percentile of individual fiber
measurements for the sample of \citet{andersen06}. We note, however, that
galaxies in \citet{andersen06} were selected in part for their face-on
inclinations, an orientation which will produce the lowest possible
measurements of velocity dispersion due to a minimal contribution from
velocity dispersions within the plane of the disk. While this may bias their results
to low $\sigma$ measurements, theirs represents the best comparison
sample currently available.

We also compare with the recent IFS studies of \citet{bello13}
and \citet{arribas14} who observe ionized gas kinematics of
LIRGS/ULIRGS at redshifts comparable to those of the
DYNAMO sample. Galaxies such as these represent the most extremely
star-forming local galaxies, and many galaxies in their sample have SFRs
comparable to those of G 4-1 and G 20-2. Similar to the results presented here,
\citet{bello13} and \citet{arribas14} find enhanced ionized gas
velocity dispersions relative to average local disk galaxies. At SFRs
in the range observed for G 4-1 and G 20-2 ($\sim$20-40 M$_{\odot}$
yr$^{-1}$), local LIRGS/ULIRGS typically have ionized gas velocity dispersions of
$\sim$60-80 km s$^{-1}$, roughly 30 km s$^{-1}$ higher than
ionized gas velocity dispersions we measure for G 4-1 and G 20-2.
This is likely due to the fact that the LIRG/ULIRG sample studied by
\citet{bello13} and \citet{arribas14} include, in addition to disks, a
significant fraction of interacting systems and mergers. Considering
only a sub-sample of non-interacting, non-AGN disk LIRGs,
\citet{arribas14} find $\sigma\simeq45$ km s$^{-1}$, comparable to the
findings presented here for DYNAMO disks.

\subsection{Stellar Kinematics and Disk Thickness}\label{section:thickness}

Both the ionized-gas and young-stellar components are observed
to have similarly high velocity dispersions \citep[compared to a velocity
dispersion of $\sim$10 km s$^{-1}$ observed for young stars in the
Milky Way,][]{wielen77,seabroke07}. We suggest that the gas
was already turbulent during the epoch in which the young, bright
stars (which dominate the observed continuum emission) were
formed. This is in contrast to a scenario in which the
stars were formed from a gas cloud with a low velocity dispersion,
which is subsequently inflated through stellar feedback to the
turbulent-gas disks currently observed. To account for this initial
turbulence, external sources of gas may be required. One possibility
is that these galaxies are undergoing large amounts of
gas accretion, which can drive turbulence through momentum or can induce
turbulence through instabilities associated with gas rich disks \citep{dekel09a,forbes13,bour14}.

What is the fate of the young stellar populations we observe in G 20-2 and G 4-1?
It is difficult to significantly alter the stellar orbits in a galaxy
short of a major merger. A violent event such as this, however, would
\textit{increase} the stellar velocity dispersions of the host
galaxies involved \citep{dasy06,schauer14}. We observe young stellar populations in G 20-2 and
G 4-1 with large velocity dispersions suggesting a thick disk of
stars. As this population ages, it will likely become a similarly
thick distribution of old stars akin to thick disks observed in
low-redshift disk galaxies. We suggest that these
galaxies contain a young, thick disk, which is in the process of forming.

We have shown in Figure~\ref{figure:expprof} that these galaxies are
well approximated by an exponential disk, allowing us to
estimate their disk thicknesses by considering the vertical
component of the stellar velocity 
dispersion.
We calculate disk scale heights using
the equation \citep{vdkAA11}:

\begin{equation}\label{equation:height}
  h_{z}=\frac{\sigma_{z}^{2}}{c \pi G \Sigma_{\mathcal{M}_{*}}}
\end{equation}

where $h_{z}$ represents the scale height, $\sigma_{z}$ is the vertical
stellar velocity dispersion, $G$ is
the gravitational constant, $\Sigma_{\mathcal{M}_{*}}$ is the
stellar mass surface density, and c is an empirical constant which varies
between 3/2 for an exponential disk and 2 for an isothermal
distribution. Here we choose a value of 3/2 which is often used in
studies of disk galaxies. We calculate our stellar mass surface density using:

\begin{equation}\label{equation:sigma}
  \Sigma_{\mathcal{M}_{*}}(r)=\frac{\mathcal{M}_{*}}{2\pi h^{2}}e^{-r/h}
\end{equation}

Where $\mathcal{M}_{*}$ is the total stellar mass, $r$ is the
radius, and $h$ is
the exponential disk scale length. We take previously reported stellar
masses from Table
\ref{table:physparam}, and we determine scale lengths from
exponential disk fits presented in Figure \ref{figure:expprof} and
also reported in Table \ref{table:physparam}.
Evaluating Equation \ref{equation:sigma} at the average radius of our
$\sigma_{d,*}$ measurements results in $\Sigma_{\mathcal{M}_{*}}$
values of 198 and 650 
M$_{\odot}$ pc$^{-2}$ for G 20-2 and G 4-1 respectively. 

What remains is to determine values of $\sigma_{z}$ from our observed
velocity dispersions.
The disk velocity dispersions reported in Section
\ref{section:skin} are line of sight values which contain a
contribution from the radial and tangential velocity dispersions
($\sigma_{R}$ and $\sigma_{\phi}$). We can recover the vertical component
of velocity dispersion using the equation \citep{shapiro03}:

\begin{equation}\label{equation:siglos}
  \sigma_{\textrm{los}}^{2}=[\sigma_{R}^{2} \sin^{2}\theta + \sigma_{\phi}^{2}
  \cos^{2} \theta] \sin^{2}i + \sigma_{z}^{2}\cos^{2}i
\end{equation}

with $i$ representing the inclination angle (determined from axis
ratios measured with GALFIT, \citealp{peng02}) and $\theta$ the
position angle from the major axis in the plane of the disk. To
isolate $\sigma_{z}$ we consider the ratios $\alpha \equiv \sigma_{z}$/$\sigma_{R}$
and $\beta \equiv \sigma_{\phi}$/$\sigma_{R}$. Combining these with Equation
\ref{equation:siglos} and taking an average over $\theta$ results in:

\begin{equation}\label{equation:sigz}
  \sigma_{z} = \frac{\sigma_{\textrm{los}}}{\cos i}
  \left(1+\frac{(\beta^{2}+1)\tan^{2}i}{2\alpha^{2}}\right)^{-1/2}
\end{equation}

From kinematics observations of the Solar Neighborhood these values
are found to be 0.5 $<$ $\alpha$ $<$ 0.6 and 0.6 $<$ $\beta$
$<$ 0.7 for the thin disk of the Milky Way. While it
is likely that observations of radially decreasing vertical velocity
dispersions will be reflected in $\alpha$ and $\beta$, radial
dependence of these values has yet to be measured
\citep{bersh10}.

Measurements of $\alpha$ and $\beta$ in external
galaxies are also limited. The best example is that of
\citet{shapiro03}, who compare integrated velocity dispersions along
the major and minor axes of 6 nearby spiral galaxies from
spectroscopic observations allowing them to measure $\alpha$
directly. This results  
in 0.5 $<$ $\alpha$ $<$ 0.8 increasing with Hubble type from Sc to
Sa. This increase, however, could simply be a result of an increased
influence of the bulge at earlier
types. Values for later types are found in the range 0.5
$<$ $\alpha$ $<$ 0.7 in agreement with measurements of
\citet{vdk99}, who measure $\alpha$ of 36 edge-on, late type galaxies (Sb - Sd)
from photometry using empirical relations of \citet{bott93}.

As discussed previously, our large values of $\sigma_{c,*}$ may
be influenced by a young galactic bulge. Thus for estimating the thickness of
our galactic disks we take $\sigma_{\textrm{los}}$ =
$\sigma_{d,*}$. For our inclination corrections we take $\alpha$ = 0.6
and $\beta$ = 0.7, but we note that these values are based on
assumptions that may be invalid for young, marginally stable disks. 

We believe it is reasonable to assume that velocity dispersions during
disk assembly 
are closer to isotropic, forcing $\alpha$ and $\beta$ to values closer
to 1. In the limit $\alpha = \beta = 1$, Equation \ref{equation:sigz} reduces to $\sigma_{z}$ =
$\sigma_{\textrm{los}}$. The large scale rotation in
these galaxies suggests that the true value of $\sigma_{z}$ will fall
somewhere between our inclination corrected value and $\sigma_{\textrm{los}}$.
We present our disk height calculations in Table \ref{table:thick}
using both $\sigma_{z}$ and $\sigma_{\textrm{los}}$ in Equation
\ref{equation:height}.

\begin{table}
  \caption{Stellar Disk Thicknesses}
  \centering
  \begin{tabular}{ c c c c c c c }
  \hline\hline
  Galaxy & $\sigma_{\textrm{los}}$ & $\sigma_{z}$ &
  $h_{z,\textrm{los}}$ & $h_{z,corr}$ \\ 
   & km s$^{-1}$ & km s$^{-1}$ & pc & pc \\
  \hline
  G 4-1 & 47.5 & 41.6 & 171 & 131 \\
  G 20-2 & 53.0 & 47.5 & 700 & 562 \\
  \hline
  \end{tabular}\label{table:thick}
\end{table}

We compare estimates of disk scale heights presented here with
\citet{yoach06} who fit two component (thin plus thick) disk models to
R-band photometry of local edge-on, late-type galaxies. 
The disk heights we estimate for G 20-2
(562-700 pc) are consistent with the high end of thin disk scale
heights and the low end for thick disks. Our scale height estimate for
G 4-1, however, is well within the thin disk regime. A caveat to our
scale height estimates is that they are very dependent on our
assumptions which are based on normal disk galaxies. These assumptions
may not be valid for turbulent, gas-rich disks such as those studied here.
We also note that
estimates of the thin disk scale height of the Milky Way range 200-300
pc \citep{vdkAA11}.  

An age dependence of stellar disk scale height is
predicted by the simulations \citet{bird13} who find that the
kinematics of a stellar population are inherited at birth. Stars
formed early are scattered by gas rich mergers \citep[or by
interactions with clumps][]{elme13b} to kinematically hot
orbits. The combination of young SSP ages and large velocity
dispersions we measure from GMOS-IFS data for galaxy G 20-2 are
consistent with the early stages of disk formation in the simulations of \citep{bird13}. 

\subsection{Stellar Populations}

The star light that we measure in 
our sample is dominated by A type stars which are responsible for
the strong Balmer absorption we observe (See Figure \ref{figure:cspecs}). From evolutionary studies of single
stellar population models it is known that the spectra of
galaxies change most significantly at early times as short lived
massive stars evolve off the main sequence. In particular we have
inferred from 
visual inspection of the P\'{E}GASE templates that the
A-star-dominated phase 
occurs approximately between 1 Myr and 1 Gyr. This qualitative assessment is consistent with the ages of our sample
from spectral fitting using a full set of P\'{E}GASE templates with
ages ranging from 1 Myr to 20 Gyr (see Section
\ref{section:tempage}). From this procedure we infer a SSP age of 500
Myr to be consistent with the stellar population which dominates the
continuum flux in our galaxies. 

It is most likely that our galaxies also host an older,
less luminous, underlying stellar population. Apart from strong Balmer
absorption lines, we also observe absorption associated with iron
which must be produced in supernovae indicating multiple generations
of stars. 
Regardless, by fitting our spectra using
spectral templates of young SSPs we are measuring the
kinematics of the young stars which are more closely associated with
the strongly emitting ionized gas.

Figure \ref{figure:cspecs} demonstrates that there is no obvious difference
between the underlying shape of the continuum spectra between the
regions of 
strongest H$\beta$ emission
and the central regions. The only major difference is the equivalent
width of H$\beta$ emission which is likely due to a local difference
in specific star formation rate. This suggests that the stellar
populations contributing the largest portion of the observed flux in
our sample are similar across the extent of our galaxies, with ages
around 500 Myr. 

\subsection{Fate of Star-Forming Clumps}\label{section:clumpfate}

We observe previously unresolved clumps of ionized gas within our
galaxies (see Figure \ref{figure:clumps}) which are completely
absent from the stellar light mapped by continuum emission.
The fate of these star-forming clumps is still uncertain. Due to their
high star formation rates it has been suggested that clumps such as these
will be dissolved by stellar feedback before they are able to migrate
into the centers of their host galaxies
\citep{murray10,genel12}. Previous studies of outflows in high
redshift clumpy galaxies have estimated clump
lifetimes in the range between 10 and 200 Myr
\citep{elme09,genzel11,newman12,wuyts12}. This is younger than the
$\chi^{2}$/DOF selected SSP age of our galaxies of 500 Myr. Figure
\ref{figure:agetst} shows however that SSP ages as young as 60 Myr
also provide reasonable kinematics fits.

We briefly mentioned in Section \ref{section:results} that we observe
evidence of a broad component in the [OIII] (5007 \AA) emission line
in galaxy G~20-2, possibly indicative of galactic scale winds. To
investigate this we create a flux-weighted mean spectrum of G~20-2
while removing emission line broadening associated with large scale rotation by
shifting each spaxel spectrally based on the ionized gas velocity map
presented in Figure \ref{figure:gaskine}. We then fit [OIII] (5007 \AA) with
a double Gaussian profile composed of a narrow and a broad component,
the latter being commonly associated with gas outflows
\citep[e.g.][]{westm12,arribas14}. We measure a broad component with a
velocity dispersion of 84 km s$^{-1}$ which is redshifted relative to
the narrow component by 5.5 km s$^{-1}$. Comparing with similar
measurement of LIRGs/ULIRGs from \citet{arribas14} we conclude that
this is evidence for weak galactic winds. This is consistent with a
scenario in which the galactic scale turbulence is not primarily
driven by star formation feedback, a result we previously inferred based
on stellar versus gas kinematics.

Reliably fitting a two component emission line profile to the Balmer
emission lines in our galaxies is hampered by the complex shape of the
corresponding strong absorption lines. For this reason, we defer
further analysis of galactic scale winds to future work which
will be based on Keck OSIRIS observations of Paschen $\alpha$.
We note here however, that the velocity dispersion of the narrow
component of [OIII] (5007 \AA) in the integrated spectra of G 20-2 is
40.0 km s$^{-1}$, consistent with mean values presented in Section
\ref{section:globalsigma}. This suggests that our results are not significantly
affected by the absence of a broad component in our emission line
fitting procedure. 

The smooth appearance of the continuum images of our galaxies
would appear to favor of short clump lifetimes, however this is not
necessarily the case. In the simulations of \citet{bour14}, large gas
outflow rates from massive ($>$10$^{8}$ M$_{\odot}$) clumps are matched
by large amounts of gas accretion. This allows clumps to survive long
enough to spiral into the center of their host galaxy while retaining
a constant mass. These simulations predict that this process should be
complete for most clumps (excluding those formed at large radii) in
less than 500 Myr. During infall the old stellar component of these clumps
will continually be stripped, possibly fueling thick disk
formation. Therefore, if these simulations are representative of the galaxies
presented here, the smooth appearance of the continuum images is not a
strong argument against long lived clumps. Such a scenario is
consistent with the theoretical work of \citet{dekel2013b}.

We show in Section \ref{section:beamsmearing} that the large velocity
dispersions we measure in the central regions of G 20-2 and G 4-1 are
not caused by beam smearing. We suggest that this is possible evidence
for a young bulge or pseudobulge formed from inspiralling clumps. From
long slit spectroscopy of the bulges of local late type galaxies,
\citet{fabri12} find velocity dispersions ranging from 50-200 km
s$^{1}$ consistent with central stellar velocity dispersions presented
here ($\sigma_{c,*}$ = 66.3 and 58.0~km~s$^{-1}$). In particular, they
find that pseudobulges (as indicated by low S\'ersic indices and
nuclear morphology that is similar to disks) have central stellar
velocity dispersion of 50-150 km s$^{-1}$ with an average of $\sim$100~km~s$^{-1}$. We note that the
largest stellar velocity dispersions we measure for single spaxels in the centers of G 
20-2 and G 4-1 are $\sim$80~km~s$^{-1}$ after correcting for beam smearing.

In both of our galaxies the
bulge-disk decomposition of the stellar surface brightness profiles
result in bulges with very low S\'ersic index ($n_b\sim 0.5-0.8$);
typically low S\'ersic indices are indicative of pseudobulges in
nearby galaxies \citep{fisher2008}. However the bulges in our galaxies
have significant differences compared to those typical of low redshift
samples. Both galaxies, particularly G20-2, are more compact than
typical of pseudobulges \citep{fisher2010}. Also it is difficult to
interpret the properties of our bulges as the central starburst in
each galaxy may affect the mass-to-light ratios. Nonetheless, a
plausible scenario may be that pseudobulges (or at least pseudobulges
in S0-Sb galaxies) are constructed through multiple mechanisms
including early build up from clumps and in more recent epochs from
secular evolution. This may be supported both by the near ubiquity of
clumpy galaxies at high redshift \citep{tacc13} and pseudobulge
galaxies at low redshift \citep{fisher2011}.  In this case, we may be
observing the early stages of pseudobulge formation in our target
galaxies.

Considering both morphological and dynamical information obtained from
GMOS-IFS data we find that galaxies G 20-2 and G 4-1 are
consistent with an early evolutionary state of massive late-type
galaxies characterized by rapid bulge growth through the infall of
gas rich clumps.

\subsection{Analogues to High Redshift Clumps?}

We observe many similarities between our galaxies and clumpy galaxies
observed at high redshift. Galaxies in the DYNAMO sample were initially selected for their extreme
H$\alpha$ emission (and thus extreme star formation) while excluding
AGN \citep{green13}. Resolved kinematic
measurements reveal a number of them to exhibit smooth rotation and
large velocity dispersions typical of high redshift samples \citep{genzel08,law09,swinbank11,wisnioski12}. Studies
at both epochs have also inferred large sSFRs based on the
Kennicut-Schmidt relation \citep{kenni98}. These findings
imply large gas fractions which have subsequently been confirmed both
at high redshift \citep{tacc10,daddi10,tacc13} and in the DYNAMO sample
(Fisher et al. submitted).

Furthermore, the GMOS observations presented here reveal two
compact disk-like galaxies from DYNAMO host clumps similar to
those observed at high redshift, providing further evidence of their
similarity to high-redshift clumpy disks.
False color images we produce highlighting
the differences between emission line regions and the stellar
continuum bear a striking resemblance to false color images of high
redshift clumpy galaxies from \citet{wuyts12} which are of similar
stellar mass to galaxies presented here. We note that this comparison may not be
entirely fair, however, as their clumpy images are produced from rest
frame ultraviolet rather than emission lines. 

Quantitatively we find, 
based on preliminary measurements of DYNAMO HST imaging, that
properties of individual DYNAMO clumps closely mirror those of clumps
at high redshift.  Both clump luminosity and size are the subject of a
forthcoming paper to which we defer detailed analysis  (Fisher et
al. in prep).  For DYNAMO galaxies G~04-1 and G~20-2 we find that
individual clumps have luminosities ranging $L_{H\alpha}\sim
10^{41}-10^{43}$~erg~s$^{-1}$ (corresponding to
$\sim1-10$~M$_{\odot}$~yr$^{-1}$). These fluxes are significantly
brighter than observations of local HII regions
\citep{wisnioski12,liverm12}, and comparable to what is observed in
clumps of high redshift galaxies \citep[e.g.][]{genzel08,wisnioski12}.

\section{Summary}\label{section:summary}

In this work we have presented the spatially-resolved
stellar-kinematics measurements of two young, star-forming 
galaxies using IFS observations with the GMOS instrument at the Gemini
Observatories. These galaxies are taken from the larger DYNAMO
sample \citep{green10,green13} of extremely H$\alpha$ luminous galaxies
selected from SDSS DR4 as possible analogues of high-redshift clumpy
disks. 

Previous studies of these objects have shown them to contain
a turbulent rotating disk of ionized gas. It was not known, however,
if the source of this  turbulence was  feedback (e.g. driven by star-formation)
or dynamical instabilities (such as caused by a gas-rich disk).
By showing that the kinematics of the ionised gas and the dominant, 
young stellar population are closely coupled we provide strong evidence
for the latter scenario. Our
findings can be summarized as follows:

\begin{itemize}
  \item For galaxy G 20-2, we confirm that the previously reported
    gas turbulence is not an artifact of beam smearing or spatial
    resolution. For galaxy
    G 4-1 we find disk velocity dispersions $\sim$20 km s$^{-1}$ lower
    than previously reported values. While beam smearing make a small
    contribution the difference is driven more by the low sensitivity
    of previous observations in the outer disk.
   
  \item In both galaxies, the rotation of the stars and ionized gas
    are closely coupled, 
    and characterized by smooth rotation and large velocity dispersions. 
    Velocity dispersions of both components are found to be $\sim10$
    km s$^{-1}$ larger
    than values measured in nearby disk galaxies ($\sim20$ km s$^{-1}$).

  \item We have determined that the stellar populations which dominate
    the optical continuum light are composed of young stars with ages
    of 60-500 Myr. Their large stellar velocity dispersion ($\sim 50$ km s$^{-1}$)
    is in marked contrast to that of the young stellar population in the disk
    of the Milky Way \citep[$\sim 5$--10 km s$^{-1}$ for $<1$ Gyr,][]{wielen77,seabroke07}.

  \item We resolve clumpy substructure in the ionized gas
    component of these galaxies that was not previously
    apparent. This supports the idea that these
    galaxies are analogous to young and clumpy disk galaxies at high redshift.

\end{itemize} 

If the star formation in these galaxies is rapidly truncated and they
continue to evolve in the absence of major mergers, they may eventually
resemble S0 galaxies or the bulges of late type galaxies we see
nearby. This is a likely outcome
as it is difficult to significantly alter the velocity profile of the
stellar component of a galaxy; the velocity dispersions of stellar
populations are effectively inherited from the gas in which they
formed. As star formation continues to gradually
deplete the gas content, the gas disk will thin and produce future
generations of stars with smaller scale heights. This scenario could also
result in galaxies similar to the Milky Way, assuming that the stellar
component we currently observe will becomes a thick disk of old
stars. If the DYNAMO sample shares the same fate as high-redshift galaxies
(as they share many physical properties) then it is possible that
high-redshift clumpy disks truly represent progenitors of disks and
S0 galaxies locally, as predicted by theory. \\
\\ \\

Support for this project is provided in part by the Victorian Department of State
Development, Business and Innovation through the Victorian
International Research Scholarship (VIRS).  We also acknowledge
support from ARC Discovery Project DP130101460. This work is based on
observations obtained at the Gemini Observatory (programs
GN-2011B-Q-54 and GS-2011B-Q-88), which is operated by
the  Association of Universities for Research in Astronomy, Inc.,
under a cooperative agreement  with the NSF on behalf of the Gemini
partnership: the National Science Foundation  (United States), the
National Research Council (Canada), CONICYT (Chile), the Australian
Research Council (Australia), Minist\'{e}rio da Ci\^{e}ncia,
Tecnologia e Inova\c{c}\~{a}o  (Brazil) and Ministerio de Ciencia,
Tecnolog\'{i}a e Innovaci\'{o}n Productiva (Argentina). 

\bibliographystyle{mn2e}

\bibliography{mn2e,alpharefs}{}

\end{document}